\documentclass[%
 reprint,
nofootinbib,
 amsmath,amssymb,
 aps
prb,
]{revtex4-1}
\pdfoutput=1
\usepackage{graphicx}
\usepackage[utf8]{inputenc}
\usepackage{flushend}
\usepackage{dcolumn}
\usepackage{bm}
\usepackage{balance}

\usepackage[colorlinks = true,
            linkcolor = blue,
            urlcolor  = blue,
            citecolor =green,
            anchorcolor = blue]{hyperref}
\usepackage{verbatim}
\usepackage{color,ulem}
\usepackage[english]{babel}

\usepackage[utf8]{inputenc}

\newcommand{\beq}{\begin{eqnarray}}
\newcommand{\eeq}{\end{eqnarray}}
\usepackage{amsmath}
\usepackage{tikz}
\usetikzlibrary{decorations.pathmorphing}
\usetikzlibrary{shapes.misc}
\tikzset{cross/.style={cross out, draw=black, minimum size=8*(#1-\pgflinewidth), inner sep=0pt, outer sep=0pt},
cross/.default={1pt}}
\usetikzlibrary{patterns,math}
\begin{document}

\title{\Large Phonon anharmonic \textcolor{black}{damping} enhances the $T_c$\\ of BCS-type superconductors}

\author{Chandan Setty}%
 \email{settychandan@gmail.com }
\affiliation{Department of Physics, University of Florida, Gainesville, Florida, USA}%

\author{Matteo Baggioli}%
 \email{matteo.baggioli@uam.es}
\affiliation{Instituto de Fisica Teorica UAM/CSIC, c/Nicolas Cabrera 13-15,
Universidad Autonoma de Madrid, Cantoblanco, 28049 Madrid, Spain.
}%

\author{Alessio Zaccone}%
 \email{alessio.zaccone@unimi.it }
\affiliation{
Department of Physics "A. Pontremoli", University of Milan, via Celoria 16, 20133 Milan, Italy.\\
Department of Chemical Engineering and Biotechnology,
University of Cambridge, Philippa Fawcett Drive, CB30AS Cambridge, U.K.\\
Cavendish Laboratory, University of Cambridge, JJ Thomson
Avenue, CB30HE Cambridge, U.K.
}

\begin{abstract}
A theory of superconductivity is presented where the effect of anharmonicity, as entailed in the acoustic, \textcolor{black}{or optical}, phonon damping, is explicitly considered in the pairing mechanism. The gap equation is solved including diffusive Akhiezer damping for longitudinal acoustic phonons \textcolor{black}{or Klemens damping for optical phonons}, with a damping coefficient which, in either case, can be directly related to the Gr{\"u}neisen parameter and hence to the anharmonic coefficients in the interatomic potential. The results show that the increase of anharmonicity has a strikingly non-monotonic effect on the critical temperature $T_{c}$. The optimal damping coefficient yielding maximum $T_c$ is set by the velocity of the bosonic mediator. This theory may open up unprecedented opportunities for material design where $T_{c}$ may be tuned via the anharmonicity of the interatomic potential, and presents implications for the superconductivity in the recently discovered hydrides, where anharmonicity is very strong and for which the \textcolor{black}{anharmonic damping} is especially relevant. 
\end{abstract}

\maketitle

\section{Introduction}
Atomic vibrations in solids are inevitably affected by the shape of the interatomic potential. For all real materials, the shape of the interatomic potential is far from being quadratic, i.e. harmonic. The intrinsic anharmonicity of solids has many well known consequences such as thermal expansion, soft modes and instabilities, sound absorption, identification of stable crystalline phases etc.~\cite{Khomskii}
A well established approach to anharmonicity is the self-consistent method introduced by Born and Hooton~\cite{Born}, leading to the concept of renormalization of phonon frequencies in the quasiharmonic or self-consistent phonon approximation, where the renormalized phonon frequencies arise from an effective vibrational dynamics within a region about equilibrium, which takes anharmonic terms of the potential into account via adjustable parameters obtained from a self-consistent solution to the many-body problem~\cite{Klein1972}.

However, the effect of anharmonicity extends to far greater areas, including electron-phonon coupling, where traditionally the effect of \textcolor{black}{anharmonic damping} has always been neglected, and where instead recent first-principle calculations demonstrate an important effect of anharmonicity on band-structure~\cite{Montserrat,Cohen}.

\textcolor{black}{In the context of high $T_c$ superconductors, the effect of anharmonic enhancement on $T_c$ has been studied in the early days following the discovery of high-$T_c$ superconductivity in the cuprates. In particular, several works by Plakida and others have studied the effect of anharmonicity on $T_c$ for the case of structurally unstable lattices or deformed lattice potentials~\cite{Plakida_1987,Zacher_1987,Plakida_1989}. Even more recent works on the high-$T_c$ hydrides~\cite{Mauri2015,Bergara2010, Mauri2013, Mauri2014, Mauri2016, Arita2016, Errea2016, Bergara2016,szcesniak2015high,errea2020quantum,camargomartnez2020higher} only take into consideration phonon energy renormalizations due to anharmonicity but neglect anharmonic damping.}

However, a  fundamental understanding of the effect of \textcolor{black}{anharmonic damping} on phonon-mediated superconductivity and e.g. on $T_{c}$ is absent due to the lack of analytical approaches to this problem. Yet, this is a fundamental issue in the context of high-T superconductors where anharmonicity becomes important due to the significant temperature values, since in general anharmonicity in solids grows roughly linear in $T$~\cite{Khomskii}. Even more urgent is the problem of the effect of anharmonicity in hydrogen-based materials, which have recorded the highest $T_{c}$ values so far: in these systems the presence of a light element such as hydrogen induces a huge anharmonicity due to the large oscillation amplitudes of the hydrogen atoms~\cite{Pickard2007,Pickard2015,Eremets2015,Pickard_review,Mauri2016, errea2020quantum}. 

Numerical studies and first-principle calculations can assess the effect of anharmonicity in an empirical way for a specific material by benchmarking against harmonic calculations, but a systematic fundamental understanding of the role of \textcolor{black}{anharmonic damping} on conventional superconductivity is missing. This would be highly beneficial to obtain system-independent guidelines to not only estimate the effect of \textcolor{black}{anharmonic damping} in general cases, but also to develop generic guidelines for material design. 
For example, by relating \textcolor{black}{anharmonic damping} to the interatomic potential it could become possible to design materials with ad-hoc or tunable electron-phonon coupling and superconducting properties.

Here we take a first step in this direction by studying the effect of \textcolor{black}{phonon Akhiezer and Klemens damping} on superconductivity beyond the quasi-harmonic approximation. We do this by explicitly taking into account the phonon damping due to anharmonicity in the mediator for the electron pairing. The theory shows that, unexpectedly, the effect of the anharmonicity (as represented by the damping coefficient) on $T_{c}$ is non-monotonic, i.e. $T_{c}$ first increases then goes through a maximum and then decreases upon increasing the anharmonic damping. This occurs because electron-phonon scattering processes involving energy-loss and energy-gain (Stokes and anti-Stokes) act constructively to increase the effective attraction driving the formation of Cooper pairs. The enhancement is most efficient for a window of critical damping parameter ($D_{max}$) set by the bosonic velocity and correlated with the Ioffe-Regel scale. Outside this window, the strength of pairing deteriorates leading a reduction in $T_c$. \textcolor{black}{These results are valid for both cases of acoustic and optical phonons, as shown in in the Appendix A.3 below.} \\

\section{The theoretical framework} 
The displacement field of an anharmonic solid obeys the following dynamical equation~\cite{Maris}:
\begin{equation}
\rho \,\frac{\partial^{2} u_{i}}{\partial t^{2}}=C^{T}_{ijkl}\, \frac{\partial^{2} u_{k}}{\partial x_{j}\partial x_{l}} - C^{T}_{ijkl}\, \alpha_{kl}\,\frac{\partial \Delta T}{\partial x_{j}} + \,\nu_{ijkl}\,\frac{\partial^{2} \dot{u}_{k}}{\partial x_{j}\partial x_{l}}
\label{DEquation}
\end{equation}
which is coupled to Fourier's law for heat transfer and to the energy balance equation for the thermal gradient $\Delta T$.
In Eq.~\eqref{DEquation} , $u_{i}$ denotes the $i$-th Cartesian component of the atomic displacement field, $C^{T}_{ijkl}$ is the isothermal elastic constant tensor, $\alpha_{kl}$ is the thermal expansion tensor, and $\nu_{ijkl}$ is the viscosity tensor. The dot indicates derivative with respect to time of the elastic field $u_{k}$ in the last dissipative term. 

For solids, where acoustic excitations can be split into longitudinal (LA) and transverse (TA), Eq.~\eqref{DEquation} can be split into two decoupled equations for LA and TA displacements, leading to the following Green's function in Fourier space ~\cite{Chaikin}:
\begin{equation}\label{GFReal}
    G_{\lambda}(\omega,q)\,=\,\frac{1}{\omega^2\,-\,\Omega_{\lambda}^{2}(q)\,+\,i\,\omega\,\Gamma_{\lambda}(q)}
\end{equation}
where $\lambda= TA, LA$ is the branch label, and $\Gamma_{\lambda}(q)=D q^{2}$ represents the \textcolor{black}{Akhiezer} damping, which 
coincides with the acoustic absorption coefficient~\cite{Maris}, while $\Omega_{\lambda}(q)=v_{\lambda}q$ is the acoustic eigenfrequency, already renormalized to account for the shift induced by anharmonicity~\cite{Dove}, with $v_{\lambda}$ the speed of sound for branch $\lambda$. 

The quadratic dependence $\Gamma_{\lambda}(q)=D q^{2}$ of the damping stems directly from the viscous term in Eq.~\eqref{DEquation} and is typical of Akhiezer damping~\cite{Akhiezer,Maris}. In particular, it has been shown \cite{Akhiezer} that $\Gamma$ takes the following general form for longitudinal excitations (see also \cite{Landau}):
\begin{equation}
\Gamma_{L}=\frac{q^{2}}{2\rho }\left[ \left(\frac{4}{3}\eta + \zeta\right) + \frac{\kappa T \alpha^{2}\rho^{2} v_{L}^{2}}{C_{p}^{2}}\left( 1 - \frac{4 v_{T}^{2}}{3 v_{L}^{2}}\right)^{2}\right].
\label{Damping}
\end{equation}
where $\eta \equiv \nu_{xyxy}$ is the shear viscosity, $\zeta $ is the bulk viscosity, $\rho$ is the solid density, $\kappa$ is the thermal conductivity, $\alpha$  is the longitudinal  thermal expansion coefficient, and $C_{p}$ is the specific heat at constant pressure.
The second term in Eq.~\eqref{Damping}, $\sim \alpha^2$, represents the phonon damping due to heat exchange between the compressed and the rarefied regions of the longitudinal wave. This second contribution, in practice, represents only a few percent of the first viscous contribution in Eq.~\eqref{Damping} and is therefore negligible.


The above derivation follows a hydrodynamic approach \cite{landau2013fluid}; by comparing with the result of a microscopic approach based on the Boltzmann transport equation for phonons, it has been shown that~\cite{Maris} 
\begin{equation}\label{DiffusionConstant-2}
D_{L} = \frac{C_{v} T \tau}{2\rho} \left(\frac{4}{3}\langle \gamma_{xy}^{2}\rangle -\langle \gamma_{xy} \rangle^{2} \right) \approx \frac{C_{v} T \tau}{2\rho} \langle \gamma_{xy}^{2}\rangle
\end{equation}
where we neglected the contribution from bulk viscosity $\zeta$, since normally $\eta \gg \zeta$. Furthermore,
$\langle...\rangle$ indicates averaging with respect to the Bose-Einstein distribution as a weight, while 
$\gamma_{xy}$ is the $xy$ component of the tensor of Gr{\"u}neisen constants. Also, $C_{v}$ is the specific heat at constant volume, while $\tau$ is the phonon life-time. Since $\tau \sim T^{-1}$ (which is an experimental observation for most solids~\cite{Boemmel,Maris}), the diffusion constant $D_L$ is independent of temperature, i.e. a well-known experimental fact~\cite{Boemmel}. 

A substantially equivalent expression for the damping of longitudinal phonons, in terms of an average Gr{\"u}neisen constant of the material $\gamma_{av}$, was derived by Boemmel and Dransfeld~\cite{Boemmel}
\begin{equation}\label{DiffusionConstant-3}
D_{L} \approx \frac{C_{v} T \tau}{2\rho} \gamma_{av}^{2}
\end{equation}
and provides a good description of the Akhiezer damping measured experimentally in quartz at $T > 60K$ \cite{Boemmel}. 

In turn, the Gr{\"u}neisen constant $\gamma$, or at least the leading term~\cite{Krivtsov2011} of $\gamma_{av}$ or $\gamma_{xy}$ above, can be directly related to the anharmonicity of the interatomic potential. For perfect crystals with pairwise nearest-neighbour interaction,  the following relation holds \cite{Krivtsov2011}
\begin{equation}
\gamma= -\frac{1}{6}\frac{V'''(a)a^{2} +2 [V''(a)a -V'(a)]}{V''(a) a+2 V'(a)}
\label{AtomicPotential}
\end{equation}
where $a$ is the equilibrium lattice spacing between nearest-neighbours, and $V'''(a)$ denotes the third derivative of the interatomic potential $V(r)$ evaluated in $r=a$. 
Hence, the phonon damping coefficient $D_{L}$ can be directly related to the anharmonicity of the interatomic potential via the Gr{\"u}neisen coefficient and Eq. \eqref{AtomicPotential}.\\

\section{Results}
Because in crystals momentum is always conserved during electron-phonon scattering events, only longitudinal phonons contribute to pairing \cite{Lee,Gorkov}, therefore we will focus on the LA phonon, $\lambda = LA$, and we will drop the $\lambda$ index in the following. 
According to Eq.~\eqref{GFReal} we thus choose a phonon propagator written in Matsubara frequency of the form 
\beq 
\Pi(i \Omega_n,\textbf{q}) = \frac{1}{v^{2}q^2 +\Omega_n^2+\Gamma(\textbf{q})\,\Omega_n},
\eeq
with $\Gamma(\textbf{q}) = D q^2$ being the Akhiezer damping discussed above, and $v$ is the phonon velocity. 
We define the Bosonic Matsubara frequency $\Omega_n = 2n\pi T$ where $n$ is an integer number and $T$ the temperature. The superconducting gap equation for a generic gap at momentum $\textbf{k}$ and Fermionic Matsubara frequency $\omega_n = (2n+1)\pi T$ takes the form (see Ref.~\cite{Carbotte2008} or ~\cite{Kleinert})
\beq \nonumber
\Delta(i\omega_n, \textbf{k}) &=& \frac{g^2}{\beta V} \sum_{\textbf{q}, \omega_m} \frac{\Delta(i\omega_m, \textbf{k}+ \textbf{q}) \Pi(\textbf{q}, i\omega_n - i \omega_m)}{\omega_m^2 + \xi_{\textbf{k} + \textbf{q}}^2 + \Delta(i\omega_m, \textbf{k}+ \textbf{q})^2}\,,\\
&&
\label{Sum-GapEqn}
\eeq
for a constant attractive interaction $g$ and volume $V$.
Here $\xi_{\textbf{k}}$ is the free electron dispersion which we choose to be quadratic with a chemical potential $\mu$. The inverse temperature is denoted by $\beta$ and we work in simplified units where twice the electron mass is set to unity.  For analytical tractability, we also choose an isotropic gap function independent of frequency, i.e., $\Delta(i\omega_m, \textbf{k}+ \textbf{q}) \equiv \Delta$. 
Converting the momentum summation into energy integral with variable $\xi$ and assuming a constant density of states, the gap equation reduces to
\beq \nonumber
1 &=& \sum_{\omega_m} \int_{-\mu}^{\infty}\frac{\lambda T d\xi}{\left[ (v^{2} - D \omega_m)(\xi + \mu)+ \omega_m^2 \right] \left[\omega_m^2 + \xi^2 + \Delta^2\right]}\\
&& 
\label{Integral-GapEqn}
\eeq
where $\lambda = N(0) g^2$ and $N(0)$ is the density of states at the Fermi level. To begin the discussion, we confine ourselves to small $D$ so that we can ignore $D\mu \ll T \sim T_c$ even though the chemical potential is allowed to be large compared to $T_c$. This implies that the linear term in $\omega_m$ can be neglected. The remaining constant $\mu v^{2}$ acts like a mass term and reduces $T_c$ for all $D$~\cite{Setty2019}. As this effect is only quantitative, this term can also be ignored, as a first approximation, without affecting the central claims of the paper.  The full effect of the chemical potential term will be included in the upcoming paragraphs. With these assumptions and using the energy integral identity $\int_{-\infty}^{\infty} \frac{d \xi}{(z \xi + s) (\xi^2 + r^2)} = \frac{\pi s}{r(s^2 + z^2 r^2)}$, we obtain
\beq \nonumber
1&=& \sum_{\omega_m}\frac{\lambda \pi T \omega_m^2}{\sqrt{\omega_m^2 + \Delta^2} \Big( \omega_m^4 + (\omega_m^2 + \Delta^2)(v^{2} - D \omega_m)^2 \Big)}.\label{pro}\\
&&
\label{MatsubaraSum-NoCP}
\eeq
 \begin{figure}[h!]
\includegraphics[width=0.8 \linewidth]{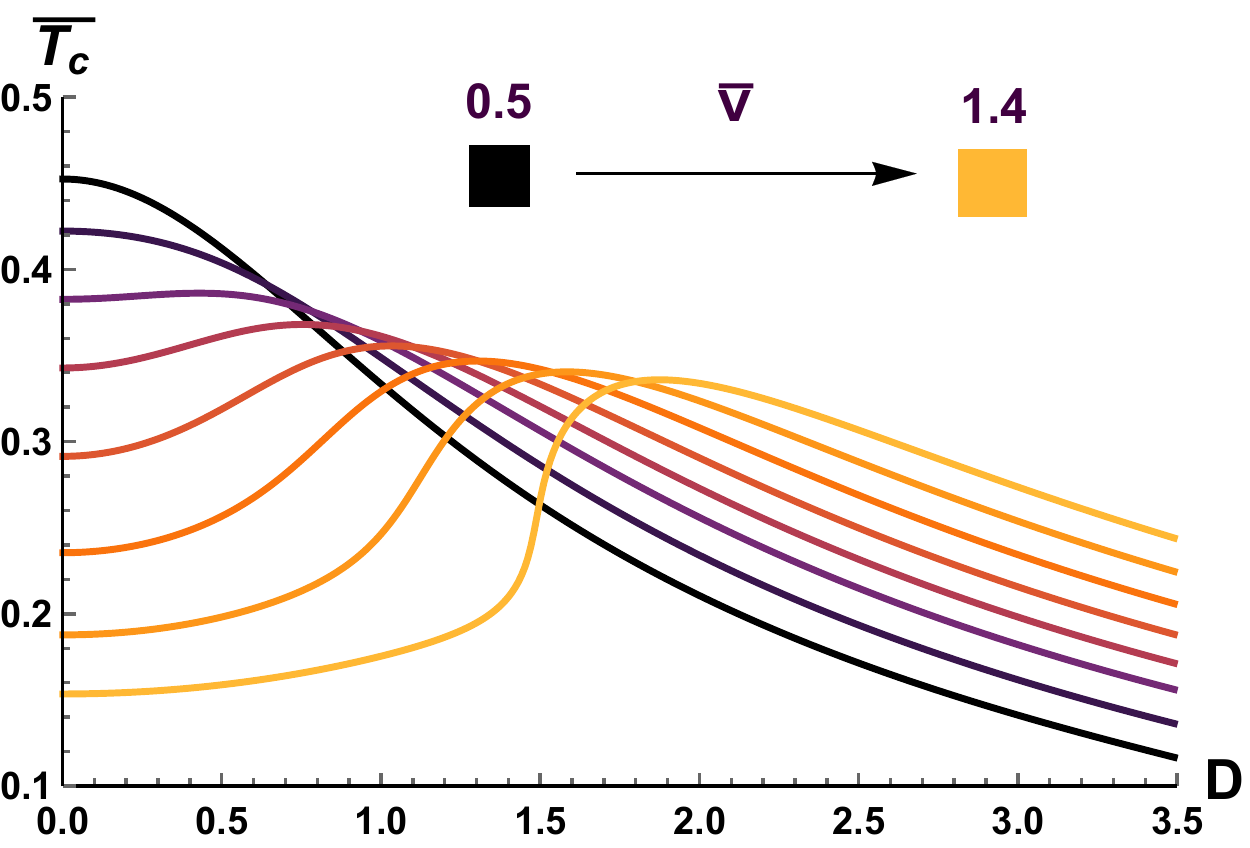}

\vspace{0.4cm}

\includegraphics[width=0.8 \linewidth]{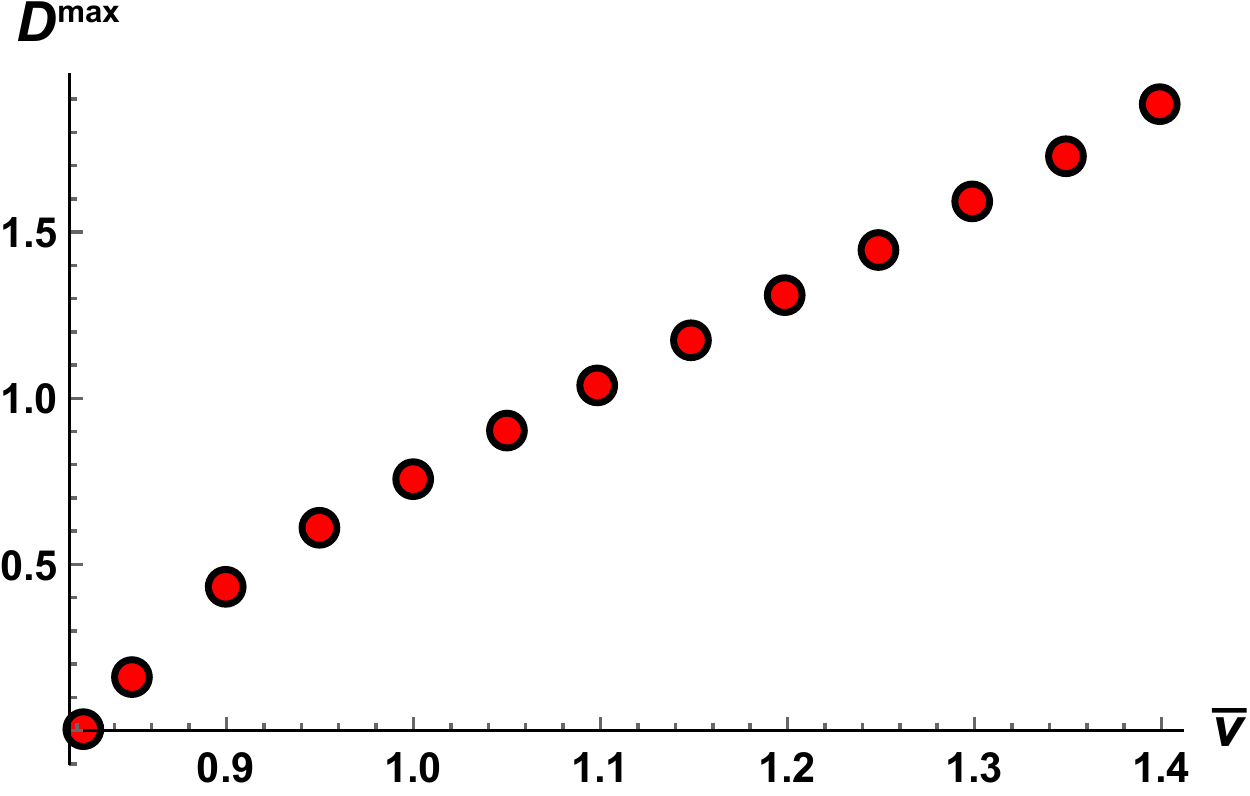}
\caption{\textbf{Top: } The dimensionless critical temperature $\bar{T}_c$ as a function of the damping constant $D$ at different dimensionless speeds $\bar{v} \in [0.5,1.4]$. \textbf{Bottom: } The position of the maximum temperature as a function of the dimensionless  longitudinal sound speed. In both plots we fixed $\bar{\mu}=0.1$.  
} \label{TcVsD} 
\end{figure}
\noindent To determine the condition for $T_c$, we set the superconducting gap $\Delta =0$. 
We can then perform the infinite sum over Matsubara frequencies (see the Appendix A.1 for more details) to obtain the simplified gap equation
\beq  \nonumber
1&=& \frac{-1}{\bar{v}^4}\Bigg[ \psi\left(\frac{1}{2}\right) + \frac{i (i + D)}{4}\psi\left(\frac{1}{2}- \frac{\bar{v}^{2}}{2 \pi \bar{T}_c (i+D)}\right)\\
&& + \frac{i (i + D)}{4}\psi\left(\frac{1}{2}+ \frac{\bar{v}^2}{2 \pi \bar{T}_c (i+D)}\right) + c.c \Bigg],
\label{NoCP}
\eeq 
\begin{figure}[hbt]
\includegraphics[width=0.8 \linewidth]{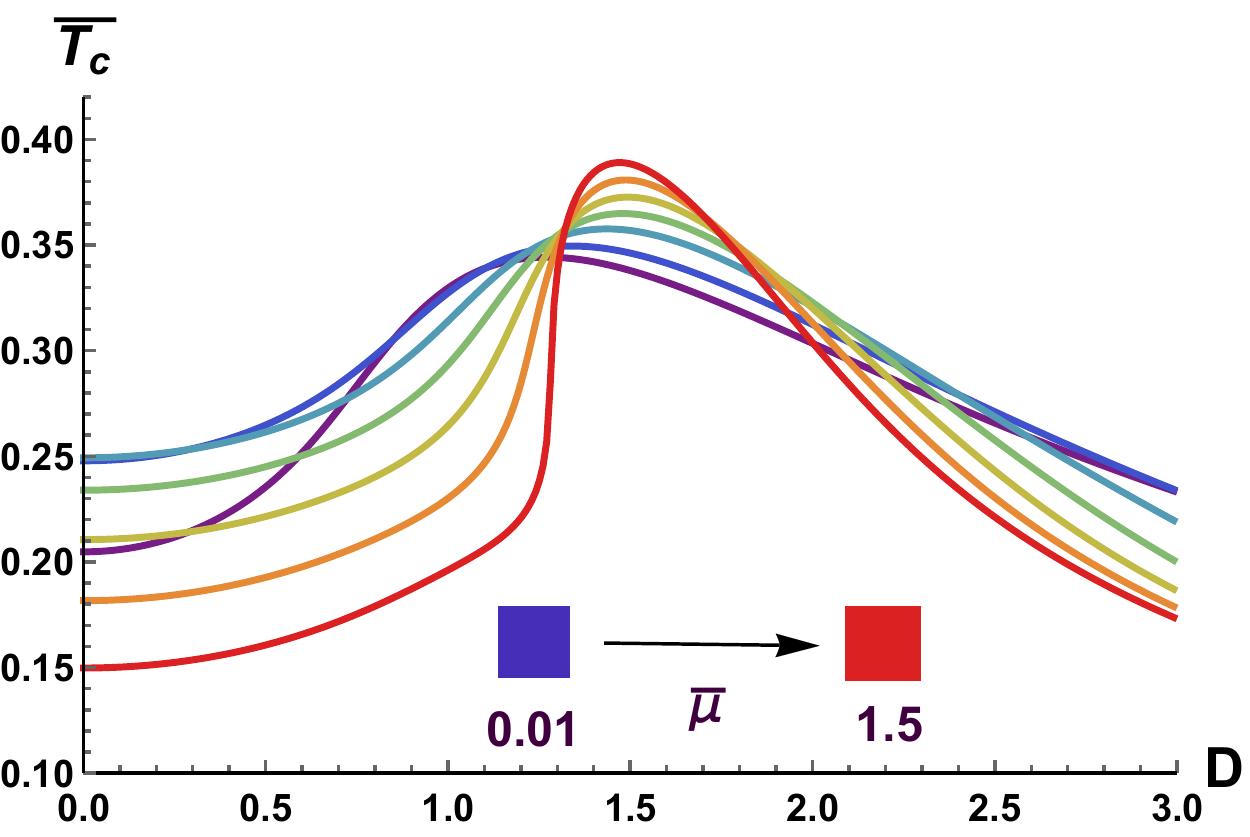}

\vspace{0.4cm}

\includegraphics[width=0.8 \linewidth]{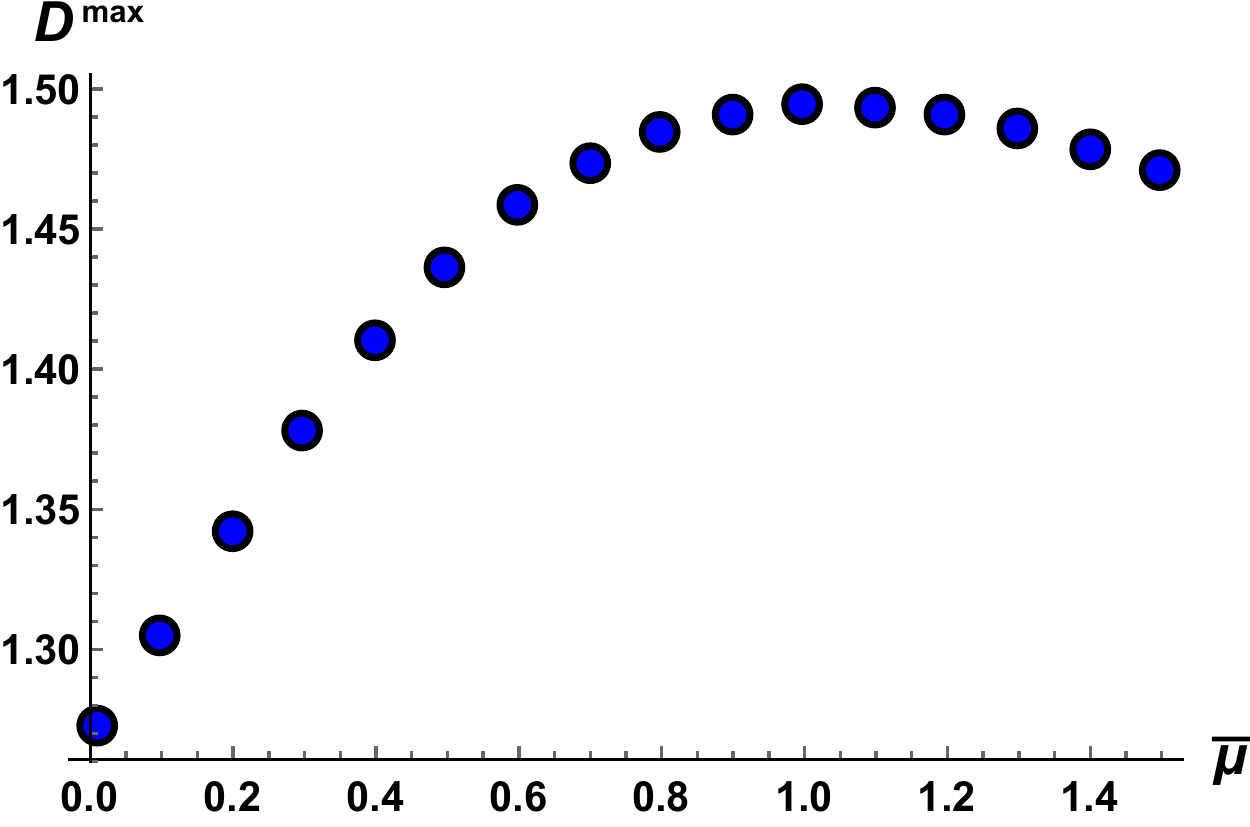}
\caption{\textbf{Top: } The dimensionless critical temperature $\bar{T}_c$ in function of the diffusion constant $D$ at different dimensionless chemical potentials $\bar{\mu} \in [0.01,1.5]$. \textbf{Bottom: } The position of the maximum temperature in function of the dimensionless chemical potential. In both plots we fixed $\bar{v}=1.2$. 
}\label{CPEffect}
\end{figure}
where, henceforth, the barred quantities are normalized by $\sqrt{\lambda}$, i.e., $\bar{v} = v/\sqrt{\lambda}$ and $\psi(x)$ is the digamma function. A solution for $\bar{T}_c$ can be obtained from Eq.~\eqref{NoCP} and is plotted in Fig.~\ref{TcVsD} (Top) as a function of the anharmonic damping parameter $D$.  The plot shows that $\bar{T}_c$ is enhanced quadratically for small $D$, reaches a maximum at an optimal anharmonicity parameter $D_{max}$ (set by the dimensionless phonon velocity $\bar{v}$), and falls off as a power law for larger $D$. The optimal parameter $D_{max}$ increases with $\bar{v}$ as shown in Fig.~\ref{TcVsD} (Bottom). In the Appendix A.2 (see also Refs.~\cite{Parshin1, Schirmacher,baggioli2019unified,Milkus,Jezowski,baggioliPRL} quoted therein), we discuss the behavior of $D_{max}$ for larger values of $\bar{v}$ where it saturates to a value $D_{max}\sim v^2/T_c$ (not shown in Fig.~\ref{TcVsD}). This condition for \textit{resonance} can be obtained from the denominator in Eq.~\ref{MatsubaraSum-NoCP}.   Note that the enhancement of the transition temperature occurs only above a critical value of the phonon velocity that is set by the interaction parameter $\sqrt{\lambda}$. The reason for the non-monotonic behavior of $\bar{T}_c$ can be understood from Eq.\eqref{pro} and the anti-symmetry in $\omega$ of the phonon damping term. Because of this property, Stokes and anti-Stokes processes ($\omega_m < \,\text{and}\, > 0$, respectively) add up constructively to increase the effective attraction driving the formation of Cooper pairs. This constructive interference grows with $D$ which gets to the numerator upon adding the two processes. Eventually, however, for sufficiently large anharmonic damping $D\gg v^2/\omega_m$, the quadratic term $\sim D^{2} \omega_m^2$ in the denominator of Eq.\eqref{pro} becomes the dominant contribution, the Stokes and anti-Stokes processes now add up in a destructive way and superconductivity gets suppressed. In the regime where $v$ is very small, the last term in the denominator of Eq.\eqref{pro} can be approximated as $(v^2-D \omega_m)^2\,\sim D^2 \omega_m^2$ and the non-monotonicity is absent even at small $D$ values (see dark lines in Fig.\ref{TcVsD}).\\

\section{Chemical potential effects} 
In the following paragraphs, we relax the assumptions made previously on the chemical potential. 
We restrict ourselves to the BCS/quasi-BCS regime where the chemical potential is positive and not below the band bottom. This assumption ignores effects where the pairing scale becomes comparable to the band-width and hence keeping the BCS-BEC cross-over regime inaccessible.  
Following the same steps of the previous section, we obtain the simplified formula 
\beq \nonumber
1 &=& \sum_{\omega_m}\frac{ \lambda \pi T_c \left(\omega_{mc}^2 + (v^{2} - D \omega_{mc}) \mu\right) |\omega_{mc}|^{-1}}{\left(\omega_{mc}^2 + (v^{2} - D \omega_{mc}) \mu\right)^2 + \omega_{mc}^2 (v^{2} - D\omega_{mc})^2}\\
&&
\label{MatsubaraSum-WithMu}
\eeq
where $\omega_{mc}$ is the Fermionic Matsubara frequency at $T=T_c$. 
After algebraic manipulations of the Matsubara sum, as shown in the Appendix A.1, the final equation for $\bar{T}_c$ with a finite chemical potential reduces to
\begin{widetext}
\beq\label{WithMu}
1 &=& \frac{1}{2\bar{v}^{2}\bar{\mu}} \Bigg[ -\psi\left(\frac{1}{2}\right) + \Bigg\{ \frac{b_+ - a^*}{2(b_+ - b_-)}\psi\left(\frac{1}{2} - b_+\right)+ \frac{a^* - b_-}{2(b_+ - b_-)}\psi\left(\frac{1}{2} - b_-\right) + c.c  \Bigg\}\Bigg] + \left[D \leftrightarrow -D\right],
\eeq
\end{widetext}
where we have the definitions $a \equiv \frac{z}{D+ i}$, $b_{\pm} \equiv \frac{z^* \pm \sqrt{z^2 + \frac{4 \bar{v}^{2} \bar{\mu}}{(2 \pi \bar{T}_c)^2}}}{2 (D-i)}$ and $z \equiv \frac{\bar{v}^{2}}{2\pi \bar{T}_c} + i D \frac{\bar{\mu}}{2\pi \bar{T}_c}$. A plot of the numerical solution for $\bar{T}_c$ versus $D$ is shown in Fig.~\ref{CPEffect} (Top). Many of the features appearing in  Fig.~\ref{TcVsD} (Top) are reproduced when the chemical potential is introduced -- a non-monotonic dependence on the anharmonicity parameter, a quadratic rise and power-law fall off for small and large $D$ respectively. This reaffirms the assumptions made on the chemical potential in deriving Eq.~\eqref{NoCP}. However, the chemical potential has an additional non-trivial effect of reducing $\bar{T}_c$ at small and large $D$, but enhances its peak value at optimal $D$. Furthermore, the $\bar{T}_c$ peak position ($D_{max}$) changes substantially for small $\bar{\mu}$ and remains virtually unchanged for larger $\bar{\mu}$.  A plot of $D_{max}$ as a function of $\bar{\mu}$ is shown in Fig.~\ref{CPEffect} (Bottom).\\

\section{Discussion} 
Much attention has been devoted to the role of disorder induced damping on superconducting $T_c$ (see~\cite{Zhu2006} and references therein); however, only a few theoretical works have examined directly the effects of damping on the superconducting properties, mostly in terms of glassiness~\cite{Seki1995, Larkin2002, Setty2019, BSZ2020}. Ref.~\cite{Larkin2002} finds an enhancement of superconducting transition driven by a spin-glass phase formed from paramagnetic spins interacting through Ruderman-Kittel-Kasuya-Yosida exchange couplings. On the other hand, Ref.~\cite{Seki1995} finds that a glassy phase leads to monotonically decreasing $T_c$ but does not take into account the role of anharmonic phonon damping explicitly. The dissipative aspect of the glass phase was considered at a phenomenological level in Ref.~\cite{Setty2019} in the context of the under-doped high-$T_c$ cuprates. While a similar non-monotonic behavior in $T_c$ is found, its mechanism does not arise from the time-reversal symmetry breaking in the dissipation term. This is reflected in the linear rise of $\bar{T}_c$ for small damping as opposed to the quadratic rise as found in this work. Furthermore, as alluded to earlier, the parameter $D$ is a characteristic of anharmonic damping and originates from the viscous damping term in Eq.~\eqref{DEquation} describing anharmonic phonons. It can be directly related to the Gr{\"u}neisen constant, which, in turn, can be determined via first-principle calculations of the inter-atomic potential through Eq.~\eqref{AtomicPotential}; therefore, this relation provides a microscopic handle for tuning $D$ giving one significant control in designing real materials. \par

\section{Conclusion}
To conclude, we have developed superconducting gap equations which account for the effect of \textcolor{black}{anharmonic damping} of phonons. The phonon viscosity parameter $D$ can be related directly to the Gr{\"u}neisen coefficient and to the shape of the interatomic potential. Upon solving the gap equation, it is found that the $T_c$ depends non-monotonically upon the anharmonic damping parameter $D$ and features a maximum as a function of $D$. The value of the critical damping parameter ($D_{max}$) around which  Cooper pairing is the strongest is set by the velocity $v$ of the phonon. Within this optimal range of damping, Stokes and anti-Stokes electron-phonon scattering processes act constructively to increase the effective coupling constant. Outside this window, the strength of pairing deteriorates leading to a reduction in $T_c$.  The prominence of the peak is enhanced when the Fermi energy is large compared to the electron-phonon coupling. Since the phonon damping corresponds to the phonon linewidth, these predictions may be further tested and investigated experimentally. \textcolor{black}{The same results (anharmonic enhancement of $T_c$ and non-monotonicity with damping) and the same resonance mechanism (this time due to Klemens damping~\cite{Klemens}) apply in the case of pairing mediated by optical phonons, as shown in the Appendix A.3 below.} Hence, the presented framework may lead to new guidelines for material design to optimize $T_c$ in conventional superconductors, including high-T hydrides. \\

\textit{Acknowledgements -- } Useful discussions with Boris Shapiro are gratefully acknowledged.
M.B. acknowledges the support of the Spanish MINECO’s ``Centro de Excelencia Severo Ochoa'' Programme under grant SEV-2012-0249. CS is supported by the U.S. DOE grant number DE-FG02-05ER46236.

\begin{appendix}
\section{Details of the derivations}
\subsection{Theoretical framework}
To obtain Eq.~\ref{Integral-GapEqn} from the gap equation (Eq.~\ref{Sum-GapEqn}; see Fig.~\ref{Feynman} for the associated self-energy diagram) we make the assumption of an isotropic gap function independent of frequency, i.e., $\Delta(i\omega_m, \textbf{k}+ \textbf{q}) \equiv \Delta$. This allows us to cancel the order parameter in the numerator on both sides of Eq.~\ref{Sum-GapEqn} and eliminate the $\omega_n$ dependence to yield
\beq  \nonumber
1 &=&  \frac{g^2}{\beta V} \sum_{\textbf{q}, \omega_m} \frac{1}{\left( (v^{2} - D \omega_m) q^2 + \omega_m^2\right) (\omega_m^2 + \xi_{\textbf{q}}^2+ \Delta^2)}.\\
&&
\eeq
We can now convert the $\textbf{q}$ momentum sum into an integral by replacing $\frac{1}{V}\sum_{\textbf{q}} \rightarrow \frac{1}{(2 \pi)^d}\int d^d\textbf{q} \rightarrow \int N(\xi)d\xi$, where $N(\xi)$ is the density of states at energy $\xi$. For quadratic bands with chemical potential $\mu$, we have $\xi_{\textbf{q}} = q^2 - \mu$ written in units stated in the main text. We now further assume a featureless density of states and approximate $N(\xi) \simeq N(0)$ as in a BCS superconductor. This is exact in two dimensions and works well when the chemical potential is far away from the band bottom in three dimensions. Defining $\lambda = g^2 N(0)$, we finally obtain Eq.~\ref{Integral-GapEqn}. \par
To obtain Eq.~\ref{NoCP} from Eq.~\ref{MatsubaraSum-NoCP}, we can simplify the Matsubara sum by summing over only positive frequencies and writing the equation for $T_c$ as 
\beq \nonumber
1 &=& \frac{\lambda}{2(2\pi T_c)^2} \sum_{m=0}^{\infty}\Bigg[ \frac{1}{x \left(x^2 + (v^{2'} + D x)^2\right)} \\
&&+  \frac{1}{x \left(x^2 + (v^{2'} - D x)^2\right)} \Bigg]
\eeq
where $x\equiv m+\frac{1}{2}$ and the primed quantities are dimensionless variables normalized by $2\pi T_c$ (i.e, $v^{2'} = v^2/2\pi T_c$). One can then use partial fractions to simplify the denominators and use the identity $\psi(z) = \lim_{k \rightarrow \infty} \left\{- \sum_{n=0}^{k-1} \frac{1}{n+z} + \ln k\right\}$. The logarithmic terms cancel to yield Eq.~\ref{NoCP}. Similarly, one can obtain Eq.~\ref{WithMu} from Eq.~\ref{MatsubaraSum-WithMu} by shifting the summation over positive frequencies and writing the equation for $T_c$ as
\begin{widetext}
\beq
1 &=& \frac{\lambda \pi T_c}{(2\pi T_c)^3} \sum_{m=0}^{\infty}\Bigg[ \frac{\left(x^2 + (v^{2'} - D x)\mu'\right)x^{-1}}{\left[\left(x^2 + (v^{2'} - D x)\mu'\right)^2 + x^2 (v^{2'} - D x)^2\right]} +  \frac{\left(x^2 + (v^{2'} + D x)\mu'\right)x^{-1}}{\left[\left(x^2 + (v^{2'} + D x)\mu'\right)^2 + x^2 (v^{2'} + D x)^2\right]} \Bigg]. \label{full}
\eeq
\end{widetext}
We again expand the summand above in partial fractions by factoring the denominators. Performing the remaining integer summations using the identity for $\psi(x)$ defined above, we obtain Eq.~\ref{WithMu}.
\begin{figure}
\centering
\begin{tikzpicture}

\node[] at (3.8,2.4) {$\Pi(\textbf{q}, i\omega_n - i\omega_m)$};
\node[] at (3.8,0.4) {$ g(\textbf{k}+\textbf{q}, i\omega_m)$};
\draw [line width=0.5mm] []
(2,0) -- (6,0);
\draw [decorate,decoration={zigzag,amplitude=.8mm,segment length=2mm}](2,0) arc (180:0:2);
\end{tikzpicture}

\caption{Feynman diagram for the anomalous self-energy. In the weak coupling BCS limit, the anomalous self-energy reduces to the gap function. The solid (zig-zag) line is the electron (boson) Green function in the superconducting state.} \label{Feynman}
\end{figure}
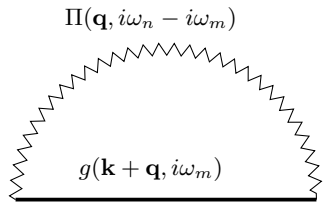\\[1cm]

\subsection{The resonance condition} 
In this paragraph, we provide more details about the resonance condition discussed in the main text. The idea is that at a specific frequency, sometimes referred to as the \textit{Ioffe-Regel frequency} \cite{Parshin1}, the boson mediator for the phonons undergoes a crossover from a ballistic propagation to a diffusive incoherent motion. More precisely, this happens at:
\begin{equation}
    \omega_{IR}\,\sim\,\frac{v^2}{\pi\,D}
\end{equation}
This value is of fundamental importance in the realm of amorphous systems, because of its correlation with the boson peak frequency, where the vibrational density of states (VDOS), normalized by the Debye law $\sim \omega^2$, displays a maximum value \cite{Schirmacher,baggioli2019unified,Milkus}. The same boson peak phenomenology, however, is also at play in strongly anharmonic crystals ~\cite{Jezowski,baggioliPRL}.

\begin{figure}
    \centering
    \includegraphics[width=0.7\linewidth]{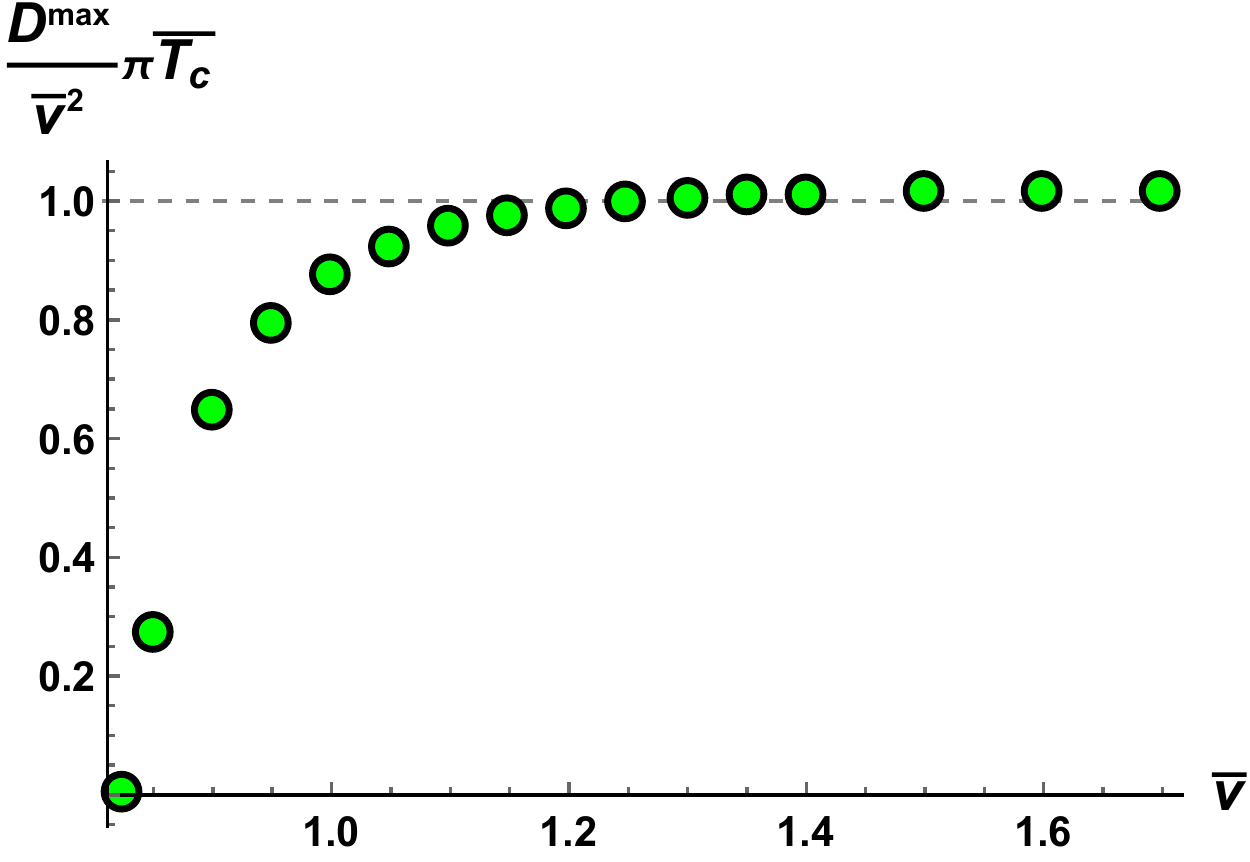}
    \caption{A validation of the resonance condition \eqref{res} using the data of fig.\ref{TcVsD}.}
    \label{figlast}
\end{figure}

Physically, this means that the density of the boson mediators is maximal around the boson peak frequency. As a consequence, one would expect the effects of the mediators to be enhanced at such energy scale. By estimating that:
\begin{equation}
    \omega_{IR}\,\sim\,T_c
\end{equation}
we arrive at the following phenomenological resonance condition:
\begin{equation}
    T_c\,\sim\,\frac{v^2}{\pi\,D_{max}}\label{res}
\end{equation}
which is quoted in the main text. Here $D_{max}$ is the value of the phonon viscosity at which $T_c$ is maximized.\\
In order to validate this expression, we plot the ratio $\pi D T_c/v^2$ in figure \ref{figlast} for the same curves shown in the main text in fig.\ref{TcVsD}. We observe, that, especially for large values of the sound speed (compared to the phonon viscosity $D$), the resonance condition \eqref{res} holds to good accuracy. This observation provides a useful correlation between the energy scale of the boson peak (induced by anharmonicity) and the maximum critical temperature that can be reached. 
\par

\subsection{Pairing mediated by anharmonic optical phonons}
In the main text we focused our attention on the case of pairing mediated by acoustic phonons, where the anharmonic damping is diffusive, $\Gamma \sim q^{2}$, according to the Akhiezer mechanism.
In this section, we consider the case of pairing mediated by optical phonons. In the case of optical phonons, the anharmonic damping is mainly related to the decay process of the optical phonon into two acoustic phonons. 
The damping coefficient $\Gamma$ is independent of $q$, in this case, and was famously calculated by Klemens using perturbation theory~\cite{Klemens}. 
As shown by Klemens, the damping parameter $\Gamma$ for optical phonons is proportional to the square of the Gr{\"u}neisen constant $\gamma$ of the material.
Hence, also in this case the $T_c$-enhancement could be tuned via the interatomic potential of the parameter through $\gamma$, in a material-by-design perspective.

\begin{figure}
    \centering
    \includegraphics[width=0.7\linewidth]{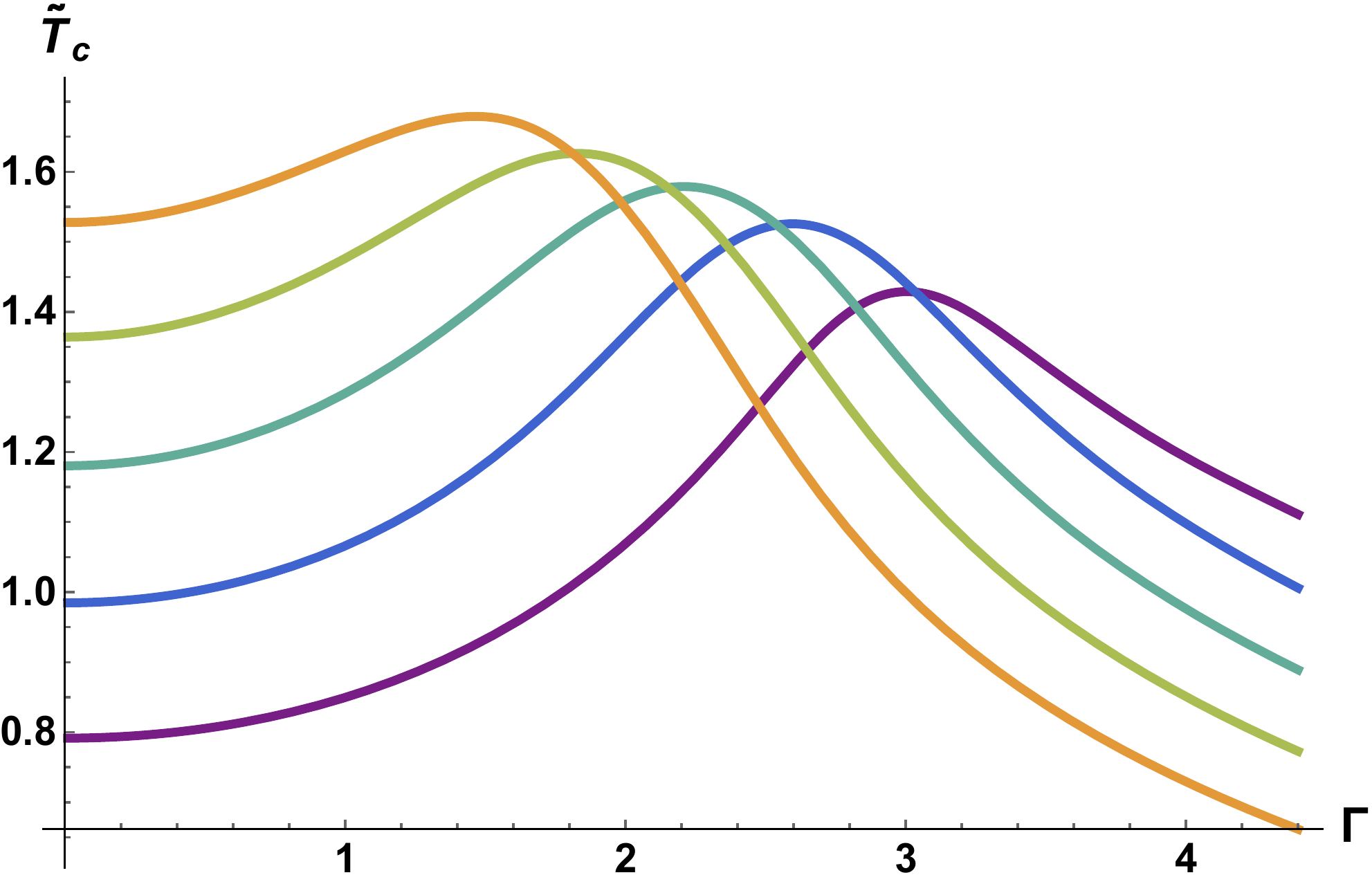}
    
    \vspace{0.2cm}
    
    \includegraphics[width=0.7\linewidth]{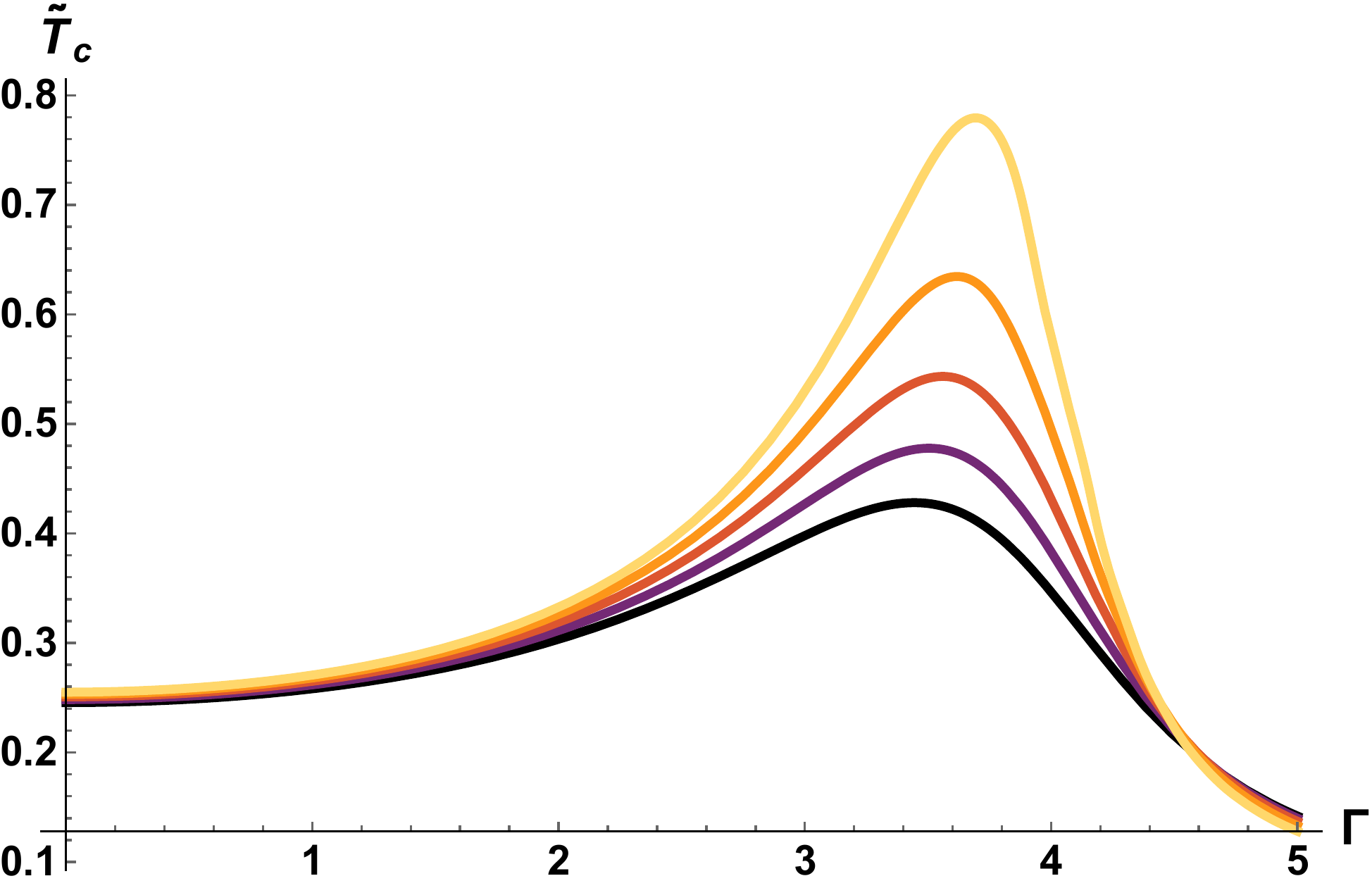}
    \caption{The dimensionless critical temperature $\tilde{T}_c\equiv 2\pi T_c/\sqrt{\lambda}$ in function of the constant damping $\Gamma$. \textbf{Top: } Increasing the mass gap of the optical mode $\omega_0^2$ from orange to purple. \textbf{Bottom:} Increasing the curvature of the optical dispersion relation $\alpha$ from yellow to black.}
    \label{figopt}
\end{figure}

Hence, we take a typical dispersion relation for optical phonons,
\begin{equation}
\Omega_{opt}(q)=\omega_0 +\,\alpha\, q^{2}
\end{equation}
with Klemens damping given a constant $\Gamma$.
We implement this model of optical phonons into the Green's function Eq. (2) of the main article, this time with damping $\Gamma = const$ independent of $q$~\cite{Klemens}, leading to the following form of the Bosonic propagator:
\begin{equation}
\Pi(i \Omega_n,\textbf{q}) = \frac{1}{\left[\omega_{0}^2 +2 \,\omega_0\,\alpha \,q^{2}+\mathcal{O}(q^4)\,\right] +\Omega_n^2-\Gamma\,\Omega_n}.
\end{equation}
Upon implementing this propagator in the theoretical framework above, we obtain the theoretical predictions for $T_c$ as a function of anharmonic damping constant $\Gamma$ for pairing mediated by optical phonons, reported in Fig.\ref{figopt} above.

These predictions align well with the effect of $T_c$-enhancement due to anharmonic damping at low damping, followed by a peak and subsequent decrease of $T_c$, that was shown in the main article for acoustic phonons. 
Also, in this case, clearly, the anharmonic damping can lead to a substantial increase of $T_c$, by at least a factor three. 
Furthermore, theory predicts that the damping-induced enhancement, and the peak, become larger upon increasing the optical phonon energy gap $\omega_0$, as shown in the top panel of Fig.\ref{figopt}.
Finally, also the curvature coefficient $\alpha$ in the optical dispersion relation has an effect on the enhancement and on the peak, they both become larger as $\alpha$ becomes smaller, hence upon approaching flat-looking optical dispersion relations, which are typically seen in DFT calculations of optical phonons in hydride materials~\cite{Pickard2015}.

\end{appendix}

\bibliographystyle{apsrev4-1}

\bibliography{anharmonicity}

\begin{thebibliography}{47}%
\makeatletter
\providecommand \@ifxundefined [1]{%
 \@ifx{#1\undefined}
}%
\providecommand \@ifnum [1]{%
 \ifnum #1\expandafter \@firstoftwo
 \else \expandafter \@secondoftwo
 \fi
}%
\providecommand \@ifx [1]{%
 \ifx #1\expandafter \@firstoftwo
 \else \expandafter \@secondoftwo
 \fi
}%
\providecommand \natexlab [1]{#1}%
\providecommand \enquote  [1]{``#1''}%
\providecommand \bibnamefont  [1]{#1}%
\providecommand \bibfnamefont [1]{#1}%
\providecommand \citenamefont [1]{#1}%
\providecommand \href@noop [0]{\@secondoftwo}%
\providecommand \href [0]{\begingroup \@sanitize@url \@href}%
\providecommand \@href[1]{\@@startlink{#1}\@@href}%
\providecommand \@@href[1]{\endgroup#1\@@endlink}%
\providecommand \@sanitize@url [0]{\catcode `\\12\catcode `\$12\catcode
  `\&12\catcode `\#12\catcode `\^12\catcode `\_12\catcode `\%12\relax}%
\providecommand \@@startlink[1]{}%
\providecommand \@@endlink[0]{}%
\providecommand \url  [0]{\begingroup\@sanitize@url \@url }%
\providecommand \@url [1]{\endgroup\@href {#1}{\urlprefix }}%
\providecommand \urlprefix  [0]{URL }%
\providecommand \Eprint [0]{\href }%
\providecommand \doibase [0]{http://dx.doi.org/}%
\providecommand \selectlanguage [0]{\@gobble}%
\providecommand \bibinfo  [0]{\@secondoftwo}%
\providecommand \bibfield  [0]{\@secondoftwo}%
\providecommand \translation [1]{[#1]}%
\providecommand \BibitemOpen [0]{}%
\providecommand \bibitemStop [0]{}%
\providecommand \bibitemNoStop [0]{.\EOS\space}%
\providecommand \EOS [0]{\spacefactor3000\relax}%
\providecommand \BibitemShut  [1]{\csname bibitem#1\endcsname}%
\let\auto@bib@innerbib\@empty
\bibitem [{\citenamefont {Khomskii}(2010)}]{Khomskii}%
  \BibitemOpen
  \bibfield  {author} {\bibinfo {author} {\bibfnamefont {D.~I.}\ \bibnamefont
  {Khomskii}},\ }\href@noop {} {\emph {\bibinfo {title} {Basic Aspects of the
  Quantum Theory of Solids}}}\ (\bibinfo  {publisher} {Cambridge University
  Press, Cambridge},\ \bibinfo {year} {2010})\BibitemShut {NoStop}%
\bibitem [{\citenamefont {Born}\ and\ \citenamefont {Hooton}(1955)}]{Born}%
  \BibitemOpen
  \bibfield  {author} {\bibinfo {author} {\bibfnamefont {M.}~\bibnamefont
  {Born}}\ and\ \bibinfo {author} {\bibfnamefont {D.~J.}\ \bibnamefont
  {Hooton}},\ }\href {\doibase 10.1007/BF01329422} {\bibfield  {journal}
  {\bibinfo  {journal} {Zeitschrift f{\"u}r Physik}\ }\textbf {\bibinfo
  {volume} {142}},\ \bibinfo {pages} {201} (\bibinfo {year}
  {1955})}\BibitemShut {NoStop}%
\bibitem [{\citenamefont {Klein}\ and\ \citenamefont
  {Horton}(1972)}]{Klein1972}%
  \BibitemOpen
  \bibfield  {author} {\bibinfo {author} {\bibfnamefont {M.~L.}\ \bibnamefont
  {Klein}}\ and\ \bibinfo {author} {\bibfnamefont {G.~K.}\ \bibnamefont
  {Horton}},\ }\href {\doibase 10.1007/BF00654839} {\bibfield  {journal}
  {\bibinfo  {journal} {Journal of Low Temperature Physics}\ }\textbf {\bibinfo
  {volume} {9}},\ \bibinfo {pages} {151} (\bibinfo {year} {1972})}\BibitemShut
  {NoStop}%
\bibitem [{\citenamefont {Monserrat}\ \emph {et~al.}(2013)\citenamefont
  {Monserrat}, \citenamefont {Drummond},\ and\ \citenamefont
  {Needs}}]{Montserrat}%
  \BibitemOpen
  \bibfield  {author} {\bibinfo {author} {\bibfnamefont {B.}~\bibnamefont
  {Monserrat}}, \bibinfo {author} {\bibfnamefont {N.~D.}\ \bibnamefont
  {Drummond}}, \ and\ \bibinfo {author} {\bibfnamefont {R.~J.}\ \bibnamefont
  {Needs}},\ }\href {\doibase 10.1103/PhysRevB.87.144302} {\bibfield  {journal}
  {\bibinfo  {journal} {Phys. Rev. B}\ }\textbf {\bibinfo {volume} {87}},\
  \bibinfo {pages} {144302} (\bibinfo {year} {2013})}\BibitemShut {NoStop}%
\bibitem [{\citenamefont {Giustino}\ \emph {et~al.}(2010)\citenamefont
  {Giustino}, \citenamefont {Louie},\ and\ \citenamefont {Cohen}}]{Cohen}%
  \BibitemOpen
  \bibfield  {author} {\bibinfo {author} {\bibfnamefont {F.}~\bibnamefont
  {Giustino}}, \bibinfo {author} {\bibfnamefont {S.~G.}\ \bibnamefont {Louie}},
  \ and\ \bibinfo {author} {\bibfnamefont {M.~L.}\ \bibnamefont {Cohen}},\
  }\href {\doibase 10.1103/PhysRevLett.105.265501} {\bibfield  {journal}
  {\bibinfo  {journal} {Phys. Rev. Lett.}\ }\textbf {\bibinfo {volume} {105}},\
  \bibinfo {pages} {265501} (\bibinfo {year} {2010})}\BibitemShut {NoStop}%
\bibitem [{\citenamefont {Plakida}\ \emph {et~al.}(1987)\citenamefont
  {Plakida}, \citenamefont {Aksenov},\ and\ \citenamefont
  {Drechsler}}]{Plakida_1987}%
  \BibitemOpen
  \bibfield  {author} {\bibinfo {author} {\bibfnamefont {N.~M.}\ \bibnamefont
  {Plakida}}, \bibinfo {author} {\bibfnamefont {V.~L.}\ \bibnamefont
  {Aksenov}}, \ and\ \bibinfo {author} {\bibfnamefont {S.~L.}\ \bibnamefont
  {Drechsler}},\ }\href {\doibase 10.1209/0295-5075/4/11/016} {\bibfield
  {journal} {\bibinfo  {journal} {Europhysics Letters ({EPL})}\ }\textbf
  {\bibinfo {volume} {4}},\ \bibinfo {pages} {1309} (\bibinfo {year}
  {1987})}\BibitemShut {NoStop}%
\bibitem [{\citenamefont {Zacher}(1987)}]{Zacher_1987}%
  \BibitemOpen
  \bibfield  {author} {\bibinfo {author} {\bibfnamefont {R.~A.}\ \bibnamefont
  {Zacher}},\ }\href {\doibase 10.1103/PhysRevB.36.7115} {\bibfield  {journal}
  {\bibinfo  {journal} {Phys. Rev. B}\ }\textbf {\bibinfo {volume} {36}},\
  \bibinfo {pages} {7115} (\bibinfo {year} {1987})}\BibitemShut {NoStop}%
\bibitem [{\citenamefont {Plakida}(1989)}]{Plakida_1989}%
  \BibitemOpen
  \bibfield  {author} {\bibinfo {author} {\bibfnamefont {N.~M.}\ \bibnamefont
  {Plakida}},\ }\href {\doibase 10.1088/0031-8949/1989/t29/013} {\bibfield
  {journal} {\bibinfo  {journal} {Physica Scripta}\ }\textbf {\bibinfo {volume}
  {T29}},\ \bibinfo {pages} {77} (\bibinfo {year} {1989})}\BibitemShut
  {NoStop}%
\bibitem [{\citenamefont {Errea}\ \emph
  {et~al.}(2015{\natexlab{a}})\citenamefont {Errea}, \citenamefont {Calandra},
  \citenamefont {Pickard}, \citenamefont {Nelson}, \citenamefont {Needs},
  \citenamefont {Li}, \citenamefont {Liu}, \citenamefont {Zhang}, \citenamefont
  {Ma},\ and\ \citenamefont {Mauri}}]{Mauri2015}%
  \BibitemOpen
  \bibfield  {author} {\bibinfo {author} {\bibfnamefont {I.}~\bibnamefont
  {Errea}}, \bibinfo {author} {\bibfnamefont {M.}~\bibnamefont {Calandra}},
  \bibinfo {author} {\bibfnamefont {C.~J.}\ \bibnamefont {Pickard}}, \bibinfo
  {author} {\bibfnamefont {J.}~\bibnamefont {Nelson}}, \bibinfo {author}
  {\bibfnamefont {R.~J.}\ \bibnamefont {Needs}}, \bibinfo {author}
  {\bibfnamefont {Y.}~\bibnamefont {Li}}, \bibinfo {author} {\bibfnamefont
  {H.}~\bibnamefont {Liu}}, \bibinfo {author} {\bibfnamefont {Y.}~\bibnamefont
  {Zhang}}, \bibinfo {author} {\bibfnamefont {Y.}~\bibnamefont {Ma}}, \ and\
  \bibinfo {author} {\bibfnamefont {F.}~\bibnamefont {Mauri}},\ }\href@noop {}
  {\bibfield  {journal} {\bibinfo  {journal} {Physical Review Letters}\
  }\textbf {\bibinfo {volume} {114}},\ \bibinfo {pages} {157004} (\bibinfo
  {year} {2015}{\natexlab{a}})}\BibitemShut {NoStop}%
\bibitem [{\citenamefont {Rousseau}\ and\ \citenamefont
  {Bergara}(2010)}]{Bergara2010}%
  \BibitemOpen
  \bibfield  {author} {\bibinfo {author} {\bibfnamefont {B.}~\bibnamefont
  {Rousseau}}\ and\ \bibinfo {author} {\bibfnamefont {A.}~\bibnamefont
  {Bergara}},\ }\href@noop {} {\bibfield  {journal} {\bibinfo  {journal}
  {Physical Review B}\ }\textbf {\bibinfo {volume} {82}},\ \bibinfo {pages}
  {104504} (\bibinfo {year} {2010})}\BibitemShut {NoStop}%
\bibitem [{\citenamefont {Errea}\ \emph {et~al.}(2013)\citenamefont {Errea},
  \citenamefont {Calandra},\ and\ \citenamefont {Mauri}}]{Mauri2013}%
  \BibitemOpen
  \bibfield  {author} {\bibinfo {author} {\bibfnamefont {I.}~\bibnamefont
  {Errea}}, \bibinfo {author} {\bibfnamefont {M.}~\bibnamefont {Calandra}}, \
  and\ \bibinfo {author} {\bibfnamefont {F.}~\bibnamefont {Mauri}},\ }\href
  {\doibase 10.1103/PhysRevLett.111.177002} {\bibfield  {journal} {\bibinfo
  {journal} {Phys. Rev. Lett.}\ }\textbf {\bibinfo {volume} {111}},\ \bibinfo
  {pages} {177002} (\bibinfo {year} {2013})}\BibitemShut {NoStop}%
\bibitem [{\citenamefont {Errea}\ \emph {et~al.}(2014)\citenamefont {Errea},
  \citenamefont {Calandra},\ and\ \citenamefont {Mauri}}]{Mauri2014}%
  \BibitemOpen
  \bibfield  {author} {\bibinfo {author} {\bibfnamefont {I.}~\bibnamefont
  {Errea}}, \bibinfo {author} {\bibfnamefont {M.}~\bibnamefont {Calandra}}, \
  and\ \bibinfo {author} {\bibfnamefont {F.}~\bibnamefont {Mauri}},\ }\href
  {\doibase 10.1103/PhysRevB.89.064302} {\bibfield  {journal} {\bibinfo
  {journal} {Phys. Rev. B}\ }\textbf {\bibinfo {volume} {89}},\ \bibinfo
  {pages} {064302} (\bibinfo {year} {2014})}\BibitemShut {NoStop}%
\bibitem [{\citenamefont {Errea}\ \emph {et~al.}(2016)\citenamefont {Errea},
  \citenamefont {Calandra}, \citenamefont {Pickard}, \citenamefont {Nelson},
  \citenamefont {Needs}, \citenamefont {Li}, \citenamefont {Liu}, \citenamefont
  {Zhang}, \citenamefont {Ma},\ and\ \citenamefont {Mauri}}]{Mauri2016}%
  \BibitemOpen
  \bibfield  {author} {\bibinfo {author} {\bibfnamefont {I.}~\bibnamefont
  {Errea}}, \bibinfo {author} {\bibfnamefont {M.}~\bibnamefont {Calandra}},
  \bibinfo {author} {\bibfnamefont {C.~J.}\ \bibnamefont {Pickard}}, \bibinfo
  {author} {\bibfnamefont {J.~R.}\ \bibnamefont {Nelson}}, \bibinfo {author}
  {\bibfnamefont {R.~J.}\ \bibnamefont {Needs}}, \bibinfo {author}
  {\bibfnamefont {Y.}~\bibnamefont {Li}}, \bibinfo {author} {\bibfnamefont
  {H.}~\bibnamefont {Liu}}, \bibinfo {author} {\bibfnamefont {Y.}~\bibnamefont
  {Zhang}}, \bibinfo {author} {\bibfnamefont {Y.}~\bibnamefont {Ma}}, \ and\
  \bibinfo {author} {\bibfnamefont {F.}~\bibnamefont {Mauri}},\ }\href@noop {}
  {\bibfield  {journal} {\bibinfo  {journal} {Nature}\ }\textbf {\bibinfo
  {volume} {532}},\ \bibinfo {pages} {81} (\bibinfo {year} {2016})}\BibitemShut
  {NoStop}%
\bibitem [{\citenamefont {Sano}\ \emph {et~al.}(2016)\citenamefont {Sano},
  \citenamefont {Koretsune}, \citenamefont {Tadano}, \citenamefont {Akashi},\
  and\ \citenamefont {Arita}}]{Arita2016}%
  \BibitemOpen
  \bibfield  {author} {\bibinfo {author} {\bibfnamefont {W.}~\bibnamefont
  {Sano}}, \bibinfo {author} {\bibfnamefont {T.}~\bibnamefont {Koretsune}},
  \bibinfo {author} {\bibfnamefont {T.}~\bibnamefont {Tadano}}, \bibinfo
  {author} {\bibfnamefont {R.}~\bibnamefont {Akashi}}, \ and\ \bibinfo {author}
  {\bibfnamefont {R.}~\bibnamefont {Arita}},\ }\href {\doibase
  10.1103/PhysRevB.93.094525} {\bibfield  {journal} {\bibinfo  {journal} {Phys.
  Rev. B}\ }\textbf {\bibinfo {volume} {93}},\ \bibinfo {pages} {094525}
  (\bibinfo {year} {2016})}\BibitemShut {NoStop}%
\bibitem [{\citenamefont {Borinaga}\ \emph
  {et~al.}(2016{\natexlab{a}})\citenamefont {Borinaga}, \citenamefont {Riego},
  \citenamefont {Leonardo}, \citenamefont {Calandra}, \citenamefont {Mauri},
  \citenamefont {Bergara},\ and\ \citenamefont {Errea}}]{Errea2016}%
  \BibitemOpen
  \bibfield  {author} {\bibinfo {author} {\bibfnamefont {M.}~\bibnamefont
  {Borinaga}}, \bibinfo {author} {\bibfnamefont {P.}~\bibnamefont {Riego}},
  \bibinfo {author} {\bibfnamefont {A.}~\bibnamefont {Leonardo}}, \bibinfo
  {author} {\bibfnamefont {M.}~\bibnamefont {Calandra}}, \bibinfo {author}
  {\bibfnamefont {F.}~\bibnamefont {Mauri}}, \bibinfo {author} {\bibfnamefont
  {A.}~\bibnamefont {Bergara}}, \ and\ \bibinfo {author} {\bibfnamefont
  {I.}~\bibnamefont {Errea}},\ }\href@noop {} {\bibfield  {journal} {\bibinfo
  {journal} {Journal of Physics: Condensed Matter}\ }\textbf {\bibinfo {volume}
  {28}},\ \bibinfo {pages} {494001} (\bibinfo {year}
  {2016}{\natexlab{a}})}\BibitemShut {NoStop}%
\bibitem [{\citenamefont {Borinaga}\ \emph
  {et~al.}(2016{\natexlab{b}})\citenamefont {Borinaga}, \citenamefont {Errea},
  \citenamefont {Calandra}, \citenamefont {Mauri},\ and\ \citenamefont
  {Bergara}}]{Bergara2016}%
  \BibitemOpen
  \bibfield  {author} {\bibinfo {author} {\bibfnamefont {M.}~\bibnamefont
  {Borinaga}}, \bibinfo {author} {\bibfnamefont {I.}~\bibnamefont {Errea}},
  \bibinfo {author} {\bibfnamefont {M.}~\bibnamefont {Calandra}}, \bibinfo
  {author} {\bibfnamefont {F.}~\bibnamefont {Mauri}}, \ and\ \bibinfo {author}
  {\bibfnamefont {A.}~\bibnamefont {Bergara}},\ }\href {\doibase
  10.1103/PhysRevB.93.174308} {\bibfield  {journal} {\bibinfo  {journal} {Phys.
  Rev. B}\ }\textbf {\bibinfo {volume} {93}},\ \bibinfo {pages} {174308}
  (\bibinfo {year} {2016}{\natexlab{b}})}\BibitemShut {NoStop}%
\bibitem [{\citenamefont {Szcze{\'s}niak}\ and\ \citenamefont
  {Zem{\l}a}(2015)}]{szcesniak2015high}%
  \BibitemOpen
  \bibfield  {author} {\bibinfo {author} {\bibfnamefont {D.}~\bibnamefont
  {Szcze{\'s}niak}}\ and\ \bibinfo {author} {\bibfnamefont {T.}~\bibnamefont
  {Zem{\l}a}},\ }\href@noop {} {\bibfield  {journal} {\bibinfo  {journal}
  {Superconductor Science and Technology}\ }\textbf {\bibinfo {volume} {28}},\
  \bibinfo {pages} {085018} (\bibinfo {year} {2015})}\BibitemShut {NoStop}%
\bibitem [{\citenamefont {Errea}\ \emph {et~al.}(2020)\citenamefont {Errea},
  \citenamefont {Belli}, \citenamefont {Monacelli}, \citenamefont {Sanna},
  \citenamefont {Koretsune}, \citenamefont {Tadano}, \citenamefont {Bianco},
  \citenamefont {Calandra}, \citenamefont {Arita}, \citenamefont {Mauri},\ and\
  \citenamefont {Flores-Livas}}]{errea2020quantum}%
  \BibitemOpen
  \bibfield  {author} {\bibinfo {author} {\bibfnamefont {I.}~\bibnamefont
  {Errea}}, \bibinfo {author} {\bibfnamefont {F.}~\bibnamefont {Belli}},
  \bibinfo {author} {\bibfnamefont {L.}~\bibnamefont {Monacelli}}, \bibinfo
  {author} {\bibfnamefont {A.}~\bibnamefont {Sanna}}, \bibinfo {author}
  {\bibfnamefont {T.}~\bibnamefont {Koretsune}}, \bibinfo {author}
  {\bibfnamefont {T.}~\bibnamefont {Tadano}}, \bibinfo {author} {\bibfnamefont
  {R.}~\bibnamefont {Bianco}}, \bibinfo {author} {\bibfnamefont
  {M.}~\bibnamefont {Calandra}}, \bibinfo {author} {\bibfnamefont
  {R.}~\bibnamefont {Arita}}, \bibinfo {author} {\bibfnamefont
  {F.}~\bibnamefont {Mauri}}, \ and\ \bibinfo {author} {\bibfnamefont {J.~A.}\
  \bibnamefont {Flores-Livas}},\ }\href {\doibase 10.1038/s41586-020-1955-z}
  {\bibfield  {journal} {\bibinfo  {journal} {Nature}\ }\textbf {\bibinfo
  {volume} {578}},\ \bibinfo {pages} {66} (\bibinfo {year} {2020})}\BibitemShut
  {NoStop}%
\bibitem [{\citenamefont {Camargo-Martínez}\ \emph {et~al.}(2020)\citenamefont
  {Camargo-Martínez}, \citenamefont {González-Pedreros},\ and\ \citenamefont
  {Mesa}}]{camargomartnez2020higher}%
  \BibitemOpen
  \bibfield  {author} {\bibinfo {author} {\bibfnamefont {J.~A.}\ \bibnamefont
  {Camargo-Martínez}}, \bibinfo {author} {\bibfnamefont {G.~I.}\ \bibnamefont
  {González-Pedreros}}, \ and\ \bibinfo {author} {\bibfnamefont
  {F.}~\bibnamefont {Mesa}},\ }\href@noop {} {\enquote {\bibinfo {title} {The
  higher superconducting transition temperature t$_c$ and the functional
  derivative of t$_c$ with $\alpha^2f(\omega)$ for electron-phonon
  superconductors},}\ } (\bibinfo {year} {2020}),\ \Eprint
  {http://arxiv.org/abs/2006.15248} {arXiv:2006.15248 [cond-mat.supr-con]}
  \BibitemShut {NoStop}%
\bibitem [{\citenamefont {Pickard}\ and\ \citenamefont
  {Needs}(2007)}]{Pickard2007}%
  \BibitemOpen
  \bibfield  {author} {\bibinfo {author} {\bibfnamefont {C.~J.}\ \bibnamefont
  {Pickard}}\ and\ \bibinfo {author} {\bibfnamefont {R.~J.}\ \bibnamefont
  {Needs}},\ }\href {\doibase 10.1038/nphys625} {\bibfield  {journal} {\bibinfo
   {journal} {Nature Physics}\ }\textbf {\bibinfo {volume} {3}},\ \bibinfo
  {pages} {473} (\bibinfo {year} {2007})}\BibitemShut {NoStop}%
\bibitem [{\citenamefont {Errea}\ \emph
  {et~al.}(2015{\natexlab{b}})\citenamefont {Errea}, \citenamefont {Calandra},
  \citenamefont {Pickard}, \citenamefont {Nelson}, \citenamefont {Needs},
  \citenamefont {Li}, \citenamefont {Liu}, \citenamefont {Zhang}, \citenamefont
  {Ma},\ and\ \citenamefont {Mauri}}]{Pickard2015}%
  \BibitemOpen
  \bibfield  {author} {\bibinfo {author} {\bibfnamefont {I.}~\bibnamefont
  {Errea}}, \bibinfo {author} {\bibfnamefont {M.}~\bibnamefont {Calandra}},
  \bibinfo {author} {\bibfnamefont {C.~J.}\ \bibnamefont {Pickard}}, \bibinfo
  {author} {\bibfnamefont {J.}~\bibnamefont {Nelson}}, \bibinfo {author}
  {\bibfnamefont {R.~J.}\ \bibnamefont {Needs}}, \bibinfo {author}
  {\bibfnamefont {Y.}~\bibnamefont {Li}}, \bibinfo {author} {\bibfnamefont
  {H.}~\bibnamefont {Liu}}, \bibinfo {author} {\bibfnamefont {Y.}~\bibnamefont
  {Zhang}}, \bibinfo {author} {\bibfnamefont {Y.}~\bibnamefont {Ma}}, \ and\
  \bibinfo {author} {\bibfnamefont {F.}~\bibnamefont {Mauri}},\ }\href
  {\doibase 10.1103/PhysRevLett.114.157004} {\bibfield  {journal} {\bibinfo
  {journal} {Phys. Rev. Lett.}\ }\textbf {\bibinfo {volume} {114}},\ \bibinfo
  {pages} {157004} (\bibinfo {year} {2015}{\natexlab{b}})}\BibitemShut
  {NoStop}%
\bibitem [{\citenamefont {Drozdov}\ \emph {et~al.}(2015)\citenamefont
  {Drozdov}, \citenamefont {Eremets}, \citenamefont {Troyan}, \citenamefont
  {Ksenofontov},\ and\ \citenamefont {Shylin}}]{Eremets2015}%
  \BibitemOpen
  \bibfield  {author} {\bibinfo {author} {\bibfnamefont {A.~P.}\ \bibnamefont
  {Drozdov}}, \bibinfo {author} {\bibfnamefont {M.~I.}\ \bibnamefont
  {Eremets}}, \bibinfo {author} {\bibfnamefont {I.~A.}\ \bibnamefont {Troyan}},
  \bibinfo {author} {\bibfnamefont {V.}~\bibnamefont {Ksenofontov}}, \ and\
  \bibinfo {author} {\bibfnamefont {S.~I.}\ \bibnamefont {Shylin}},\ }\href
  {\doibase 10.1038/nature14964} {\bibfield  {journal} {\bibinfo  {journal}
  {Nature}\ }\textbf {\bibinfo {volume} {525}},\ \bibinfo {pages} {73}
  (\bibinfo {year} {2015})}\BibitemShut {NoStop}%
\bibitem [{\citenamefont {Pickard}\ \emph {et~al.}(2020)\citenamefont
  {Pickard}, \citenamefont {Errea},\ and\ \citenamefont
  {Eremets}}]{Pickard_review}%
  \BibitemOpen
  \bibfield  {author} {\bibinfo {author} {\bibfnamefont {C.~J.}\ \bibnamefont
  {Pickard}}, \bibinfo {author} {\bibfnamefont {I.}~\bibnamefont {Errea}}, \
  and\ \bibinfo {author} {\bibfnamefont {M.~I.}\ \bibnamefont {Eremets}},\
  }\href {\doibase 10.1146/annurev-conmatphys-031218-013413} {\bibfield
  {journal} {\bibinfo  {journal} {Annual Review of Condensed Matter Physics}\
  }\textbf {\bibinfo {volume} {11}},\ \bibinfo {pages} {null} (\bibinfo {year}
  {2020})}\BibitemShut {NoStop}%
\bibitem [{\citenamefont {Maris}(1971)}]{Maris}%
  \BibitemOpen
  \bibfield  {author} {\bibinfo {author} {\bibfnamefont {H.~J.}\ \bibnamefont
  {Maris}},\ }\href@noop {} {\emph {\bibinfo {title} {Physical Acoustics, vol.
  VIII}}},\ edited by\ \bibinfo {editor} {\bibfnamefont {W.}~\bibnamefont
  {Mason}}\ and\ \bibinfo {editor} {\bibfnamefont {R.}~\bibnamefont
  {Thurston}}\ (\bibinfo  {publisher} {Academic Press, London},\ \bibinfo
  {year} {1971})\BibitemShut {NoStop}%
\bibitem [{\citenamefont {Chaikin}\ and\ \citenamefont
  {Lubensky}(1995)}]{Chaikin}%
  \BibitemOpen
  \bibfield  {author} {\bibinfo {author} {\bibfnamefont {P.~M.}\ \bibnamefont
  {Chaikin}}\ and\ \bibinfo {author} {\bibfnamefont {T.~C.}\ \bibnamefont
  {Lubensky}},\ }\href {\doibase 10.1017/CBO9780511813467} {\emph {\bibinfo
  {title} {Principles of Condensed Matter Physics}}}\ (\bibinfo  {publisher}
  {Cambridge University Press},\ \bibinfo {year} {1995})\BibitemShut {NoStop}%
\bibitem [{\citenamefont {Dove}(1993)}]{Dove}%
  \BibitemOpen
  \bibfield  {author} {\bibinfo {author} {\bibfnamefont {M.~T.}\ \bibnamefont
  {Dove}},\ }\href@noop {} {\emph {\bibinfo {title} {Introduction to lattice
  dynamics}}}\ (\bibinfo  {publisher} {Cambridge University Press},\ \bibinfo
  {year} {1993})\BibitemShut {NoStop}%
\bibitem [{\citenamefont {Akhiezer}(1939)}]{Akhiezer}%
  \BibitemOpen
  \bibfield  {author} {\bibinfo {author} {\bibfnamefont {A.~I.}\ \bibnamefont
  {Akhiezer}},\ }\href@noop {} {\bibfield  {journal} {\bibinfo  {journal} {J.
  Phys. (Moscow)}\ }\textbf {\bibinfo {volume} {1}},\ \bibinfo {pages} {277}
  (\bibinfo {year} {1939})}\BibitemShut {NoStop}%
\bibitem [{\citenamefont {Landau}\ and\ \citenamefont
  {Lifshitz}(1986)}]{Landau}%
  \BibitemOpen
  \bibfield  {author} {\bibinfo {author} {\bibfnamefont {L.~D.}\ \bibnamefont
  {Landau}}\ and\ \bibinfo {author} {\bibfnamefont {E.~M.}\ \bibnamefont
  {Lifshitz}},\ }\href@noop {} {\emph {\bibinfo {title} {Theory of
  Elasticity}}}\ (\bibinfo  {publisher} {Butterworth-Heinemann, Oxford},\
  \bibinfo {year} {1986})\BibitemShut {NoStop}%
\bibitem [{\citenamefont {Landau}\ and\ \citenamefont
  {Lifshitz}(2013)}]{landau2013fluid}%
  \BibitemOpen
  \bibfield  {author} {\bibinfo {author} {\bibfnamefont {L.}~\bibnamefont
  {Landau}}\ and\ \bibinfo {author} {\bibfnamefont {E.}~\bibnamefont
  {Lifshitz}},\ }\href {https://books.google.es/books?id=CeBbAwAAQBAJ} {\emph
  {\bibinfo {title} {Fluid Mechanics}}},\ \bibinfo {number} {v. 6}\ (\bibinfo
  {publisher} {Elsevier Science},\ \bibinfo {year} {2013})\BibitemShut
  {NoStop}%
\bibitem [{\citenamefont {B\"ommel}\ and\ \citenamefont
  {Dransfeld}(1960)}]{Boemmel}%
  \BibitemOpen
  \bibfield  {author} {\bibinfo {author} {\bibfnamefont {H.~E.}\ \bibnamefont
  {B\"ommel}}\ and\ \bibinfo {author} {\bibfnamefont {K.}~\bibnamefont
  {Dransfeld}},\ }\href {\doibase 10.1103/PhysRev.117.1245} {\bibfield
  {journal} {\bibinfo  {journal} {Phys. Rev.}\ }\textbf {\bibinfo {volume}
  {117}},\ \bibinfo {pages} {1245} (\bibinfo {year} {1960})}\BibitemShut
  {NoStop}%
\bibitem [{\citenamefont {Krivtsov}\ and\ \citenamefont
  {Kuz'kin}(2011)}]{Krivtsov2011}%
  \BibitemOpen
  \bibfield  {author} {\bibinfo {author} {\bibfnamefont {A.~M.}\ \bibnamefont
  {Krivtsov}}\ and\ \bibinfo {author} {\bibfnamefont {V.~A.}\ \bibnamefont
  {Kuz'kin}},\ }\href {\doibase 10.3103/S002565441103006X} {\bibfield
  {journal} {\bibinfo  {journal} {Mechanics of Solids}\ }\textbf {\bibinfo
  {volume} {46}},\ \bibinfo {pages} {387} (\bibinfo {year} {2011})}\BibitemShut
  {NoStop}%
\bibitem [{\citenamefont {Ruhman}\ and\ \citenamefont {Lee}(2019)}]{Lee}%
  \BibitemOpen
  \bibfield  {author} {\bibinfo {author} {\bibfnamefont {J.}~\bibnamefont
  {Ruhman}}\ and\ \bibinfo {author} {\bibfnamefont {P.~A.}\ \bibnamefont
  {Lee}},\ }\href {\doibase 10.1103/PhysRevB.100.226501} {\bibfield  {journal}
  {\bibinfo  {journal} {Phys. Rev. B}\ }\textbf {\bibinfo {volume} {100}},\
  \bibinfo {pages} {226501} (\bibinfo {year} {2019})}\BibitemShut {NoStop}%
\bibitem [{\citenamefont {Gor{\textquoteright}kov}(2016)}]{Gorkov}%
  \BibitemOpen
  \bibfield  {author} {\bibinfo {author} {\bibfnamefont {L.~P.}\ \bibnamefont
  {Gor{\textquoteright}kov}},\ }\href {\doibase 10.1073/pnas.1604145113}
  {\bibfield  {journal} {\bibinfo  {journal} {Proceedings of the National
  Academy of Sciences}\ }\textbf {\bibinfo {volume} {113}},\ \bibinfo {pages}
  {4646} (\bibinfo {year} {2016})}\BibitemShut {NoStop}%
\bibitem [{\citenamefont {Marsiglio}\ and\ \citenamefont
  {Carbotte}(2008)}]{Carbotte2008}%
  \BibitemOpen
  \bibfield  {author} {\bibinfo {author} {\bibfnamefont {F.}~\bibnamefont
  {Marsiglio}}\ and\ \bibinfo {author} {\bibfnamefont {J.}~\bibnamefont
  {Carbotte}},\ }in\ \href@noop {} {\emph {\bibinfo {booktitle}
  {Superconductivity}}}\ (\bibinfo  {publisher} {Springer},\ \bibinfo {year}
  {2008})\ pp.\ \bibinfo {pages} {73--162}\BibitemShut {NoStop}%
\bibitem [{\citenamefont {Kleinert}(2018)}]{Kleinert}%
  \BibitemOpen
  \bibfield  {author} {\bibinfo {author} {\bibfnamefont {H.}~\bibnamefont
  {Kleinert}},\ }\href {\doibase 10.1142/10545} {\emph {\bibinfo {title}
  {Collective classical and quantum fields}}}\ (\bibinfo  {publisher} {World
  Scientific, Singapore},\ \bibinfo {year} {2018})\BibitemShut {NoStop}%
\bibitem [{\citenamefont {Setty}(2019)}]{Setty2019}%
  \BibitemOpen
  \bibfield  {author} {\bibinfo {author} {\bibfnamefont {C.}~\bibnamefont
  {Setty}},\ }\href {\doibase 10.1103/PhysRevB.99.144523} {\bibfield  {journal}
  {\bibinfo  {journal} {Phys. Rev. B}\ }\textbf {\bibinfo {volume} {99}},\
  \bibinfo {pages} {144523} (\bibinfo {year} {2019})}\BibitemShut {NoStop}%
\bibitem [{\citenamefont {Beltukov}\ \emph {et~al.}(2013)\citenamefont
  {Beltukov}, \citenamefont {Kozub},\ and\ \citenamefont {Parshin}}]{Parshin1}%
  \BibitemOpen
  \bibfield  {author} {\bibinfo {author} {\bibfnamefont {Y.~M.}\ \bibnamefont
  {Beltukov}}, \bibinfo {author} {\bibfnamefont {V.~I.}\ \bibnamefont {Kozub}},
  \ and\ \bibinfo {author} {\bibfnamefont {D.~A.}\ \bibnamefont {Parshin}},\
  }\href {\doibase 10.1103/PhysRevB.87.134203} {\bibfield  {journal} {\bibinfo
  {journal} {Phys. Rev. B}\ }\textbf {\bibinfo {volume} {87}},\ \bibinfo
  {pages} {134203} (\bibinfo {year} {2013})}\BibitemShut {NoStop}%
\bibitem [{\citenamefont {Marruzzo}\ \emph {et~al.}(2013)\citenamefont
  {Marruzzo}, \citenamefont {Schirmacher}, \citenamefont {Fratalocchi},\ and\
  \citenamefont {Ruocco}}]{Schirmacher}%
  \BibitemOpen
  \bibfield  {author} {\bibinfo {author} {\bibfnamefont {A.}~\bibnamefont
  {Marruzzo}}, \bibinfo {author} {\bibfnamefont {W.}~\bibnamefont
  {Schirmacher}}, \bibinfo {author} {\bibfnamefont {A.}~\bibnamefont
  {Fratalocchi}}, \ and\ \bibinfo {author} {\bibfnamefont {G.}~\bibnamefont
  {Ruocco}},\ }\href {https://doi.org/10.1038/srep01407} {\bibfield  {journal}
  {\bibinfo  {journal} {Scientific Reports}\ }\textbf {\bibinfo {volume} {3}},\
  \bibinfo {pages} {1407 EP } (\bibinfo {year} {2013})},\ \bibinfo {note}
  {article}\BibitemShut {NoStop}%
\bibitem [{\citenamefont {Baggioli}\ and\ \citenamefont
  {Zaccone}(2020)}]{baggioli2019unified}%
  \BibitemOpen
  \bibfield  {author} {\bibinfo {author} {\bibfnamefont {M.}~\bibnamefont
  {Baggioli}}\ and\ \bibinfo {author} {\bibfnamefont {A.}~\bibnamefont
  {Zaccone}},\ }\href {\doibase 10.1103/PhysRevResearch.2.013267} {\bibfield
  {journal} {\bibinfo  {journal} {Phys. Rev. Research}\ }\textbf {\bibinfo
  {volume} {2}},\ \bibinfo {pages} {013267} (\bibinfo {year}
  {2020})}\BibitemShut {NoStop}%
\bibitem [{\citenamefont {Milkus}\ and\ \citenamefont
  {Zaccone}(2016)}]{Milkus}%
  \BibitemOpen
  \bibfield  {author} {\bibinfo {author} {\bibfnamefont {R.}~\bibnamefont
  {Milkus}}\ and\ \bibinfo {author} {\bibfnamefont {A.}~\bibnamefont
  {Zaccone}},\ }\href {\doibase 10.1103/PhysRevB.93.094204} {\bibfield
  {journal} {\bibinfo  {journal} {Phys. Rev. B}\ }\textbf {\bibinfo {volume}
  {93}},\ \bibinfo {pages} {094204} (\bibinfo {year} {2016})}\BibitemShut
  {NoStop}%
\bibitem [{\citenamefont {Je\ifmmode~\dot{z}\else \.{z}\fi{}owski}\ \emph
  {et~al.}(2018)\citenamefont {Je\ifmmode~\dot{z}\else \.{z}\fi{}owski},
  \citenamefont {Strzhemechny}, \citenamefont {Krivchikov}, \citenamefont
  {Davydova}, \citenamefont {Szewczyk}, \citenamefont {Stepanian},
  \citenamefont {Buravtseva},\ and\ \citenamefont {Romantsova}}]{Jezowski}%
  \BibitemOpen
  \bibfield  {author} {\bibinfo {author} {\bibfnamefont {A.}~\bibnamefont
  {Je\ifmmode~\dot{z}\else \.{z}\fi{}owski}}, \bibinfo {author} {\bibfnamefont
  {M.~A.}\ \bibnamefont {Strzhemechny}}, \bibinfo {author} {\bibfnamefont
  {A.~I.}\ \bibnamefont {Krivchikov}}, \bibinfo {author} {\bibfnamefont
  {N.~A.}\ \bibnamefont {Davydova}}, \bibinfo {author} {\bibfnamefont
  {D.}~\bibnamefont {Szewczyk}}, \bibinfo {author} {\bibfnamefont {S.~G.}\
  \bibnamefont {Stepanian}}, \bibinfo {author} {\bibfnamefont {L.~M.}\
  \bibnamefont {Buravtseva}}, \ and\ \bibinfo {author} {\bibfnamefont {O.~O.}\
  \bibnamefont {Romantsova}},\ }\href {\doibase 10.1103/PhysRevB.97.201201}
  {\bibfield  {journal} {\bibinfo  {journal} {Phys. Rev. B}\ }\textbf {\bibinfo
  {volume} {97}},\ \bibinfo {pages} {201201} (\bibinfo {year}
  {2018})}\BibitemShut {NoStop}%
\bibitem [{\citenamefont {Baggioli}\ and\ \citenamefont
  {Zaccone}(2019)}]{baggioliPRL}%
  \BibitemOpen
  \bibfield  {author} {\bibinfo {author} {\bibfnamefont {M.}~\bibnamefont
  {Baggioli}}\ and\ \bibinfo {author} {\bibfnamefont {A.}~\bibnamefont
  {Zaccone}},\ }\href {\doibase 10.1103/PhysRevLett.122.145501} {\bibfield
  {journal} {\bibinfo  {journal} {Phys. Rev. Lett.}\ }\textbf {\bibinfo
  {volume} {122}},\ \bibinfo {pages} {145501} (\bibinfo {year}
  {2019})}\BibitemShut {NoStop}%
\bibitem [{\citenamefont {Balatsky}\ \emph {et~al.}(2006)\citenamefont
  {Balatsky}, \citenamefont {Vekhter},\ and\ \citenamefont {Zhu}}]{Zhu2006}%
  \BibitemOpen
  \bibfield  {author} {\bibinfo {author} {\bibfnamefont {A.~V.}\ \bibnamefont
  {Balatsky}}, \bibinfo {author} {\bibfnamefont {I.}~\bibnamefont {Vekhter}}, \
  and\ \bibinfo {author} {\bibfnamefont {J.-X.}\ \bibnamefont {Zhu}},\
  }\href@noop {} {\bibfield  {journal} {\bibinfo  {journal} {Reviews of Modern
  Physics}\ }\textbf {\bibinfo {volume} {78}},\ \bibinfo {pages} {373}
  (\bibinfo {year} {2006})}\BibitemShut {NoStop}%
\bibitem [{\citenamefont {Seki}(1995)}]{Seki1995}%
  \BibitemOpen
  \bibfield  {author} {\bibinfo {author} {\bibfnamefont {S.}~\bibnamefont
  {Seki}},\ }\href@noop {} {\bibfield  {journal} {\bibinfo  {journal} {Progress
  of Theoretical Physics}\ }\textbf {\bibinfo {volume} {94}},\ \bibinfo {pages}
  {181} (\bibinfo {year} {1995})}\BibitemShut {NoStop}%
\bibitem [{\citenamefont {Galitski}\ and\ \citenamefont
  {Larkin}(2002)}]{Larkin2002}%
  \BibitemOpen
  \bibfield  {author} {\bibinfo {author} {\bibfnamefont {V.}~\bibnamefont
  {Galitski}}\ and\ \bibinfo {author} {\bibfnamefont {A.}~\bibnamefont
  {Larkin}},\ }\href@noop {} {\bibfield  {journal} {\bibinfo  {journal}
  {Physical Review B}\ }\textbf {\bibinfo {volume} {66}},\ \bibinfo {pages}
  {064526} (\bibinfo {year} {2002})}\BibitemShut {NoStop}%
\bibitem [{\citenamefont {Baggioli}\ \emph {et~al.}(2020)\citenamefont
  {Baggioli}, \citenamefont {Setty},\ and\ \citenamefont {Zaccone}}]{BSZ2020}%
  \BibitemOpen
  \bibfield  {author} {\bibinfo {author} {\bibfnamefont {M.}~\bibnamefont
  {Baggioli}}, \bibinfo {author} {\bibfnamefont {C.}~\bibnamefont {Setty}}, \
  and\ \bibinfo {author} {\bibfnamefont {A.}~\bibnamefont {Zaccone}},\ }\href
  {\doibase 10.1103/PhysRevB.101.214502} {\bibfield  {journal} {\bibinfo
  {journal} {Phys. Rev. B}\ }\textbf {\bibinfo {volume} {101}},\ \bibinfo
  {pages} {214502} (\bibinfo {year} {2020})}\BibitemShut {NoStop}%
\bibitem [{\citenamefont {Klemens}(1966)}]{Klemens}%
  \BibitemOpen
  \bibfield  {author} {\bibinfo {author} {\bibfnamefont {P.~G.}\ \bibnamefont
  {Klemens}},\ }\href {\doibase 10.1103/PhysRev.148.845} {\bibfield  {journal}
  {\bibinfo  {journal} {Phys. Rev.}\ }\textbf {\bibinfo {volume} {148}},\
  \bibinfo {pages} {845} (\bibinfo {year} {1966})}\BibitemShut {NoStop}%
\end{thebibliography}%

\end{document}